\documentclass[twocolumn, switch]{article} % Method A for two-column formatting

\usepackage{preprint}

\usepackage{amsmath, amsthm, amssymb, amsfonts}

\usepackage{multirow}
\usepackage{threeparttable}
\usepackage{subcaption}
\usepackage{booktabs,caption}
\usepackage{threeparttable}
\usepackage{enumerate}

\newcommand{\addtabletext}[1]{{\setlength{\leftskip}{9pt}\fontsize{7}{9}\selectfont#1}}

\DeclareMathOperator*{\argmax}{arg\,max}

\usepackage[numbers,square]{natbib}
\usepackage[utf8]{inputenc}	% allow utf-8 input
\usepackage[T1]{fontenc}	% use 8-bit T1 fonts
\usepackage{xcolor}		% colors for hyperlinks
\usepackage[colorlinks = true,
            linkcolor = purple,
            urlcolor  = blue,
            citecolor = cyan,
            anchorcolor = black]{hyperref}	% Color links to references, figures, etc.
\usepackage{booktabs} 		% professional-quality tables
\usepackage{nicefrac}		% compact symbols for 1/2, etc.
\usepackage{microtype}		% microtypography
\usepackage{lineno}		% Line numbers
\usepackage{float}			% Allows for figures within multicol

\usepackage{lipsum}		%  Filler text

\usepackage{newfloat}
\DeclareFloatingEnvironment[name={Supplementary Figure}]{suppfigure}
\usepackage{sidecap}
\sidecaptionvpos{figure}{c}

\usepackage{titlesec}
\titlespacing\section{0pt}{12pt plus 3pt minus 3pt}{1pt plus 1pt minus 1pt}
\titlespacing\subsection{0pt}{10pt plus 3pt minus 3pt}{1pt plus 1pt minus 1pt}
\titlespacing\subsubsection{0pt}{8pt plus 3pt minus 3pt}{1pt plus 1pt minus 1pt}

\usepackage{array}
\usepackage{multirow}
\usepackage[]{threeparttable}

\newcolumntype{P}[1]{>{\centering\arraybackslash}p{#1}}

\title{Multiscale Flow for Robust and Optimal Cosmological Analysis}

\usepackage{authblk}

\author[a,b\thanks{\tt{biwei@berkeley.edu}}]{Biwei Dai}
\author[a,b,c]{Uro{\v s} Seljak}
\affil[a]{Berkeley Center for Cosmological Physics and Department of Physics, University of California, Berkeley, CA 94720, USA}
\affil[b]{Physics Division, Lawrence Berkeley National Lab, 1 Cyclotron Road, Berkeley, CA 94720, USA}
\affil[c]{Department of Astronomy, University of California, Berkeley, CA 94720, USA}
\begin{document}

\twocolumn[ % Method A for two-column formatting
  \begin{@twocolumnfalse} % Method A for two-column formatting
  
\maketitle

\begin{abstract}
We propose Multiscale Flow, a generative Normalizing Flow that creates samples and models the field-level likelihood of two-dimensional cosmological data such as weak lensing.
%, thus enabling Simulation Based Likelihood Inference. 
Multiscale Flow uses hierarchical decomposition of cosmological fields via a wavelet basis, and then models different wavelet components separately as Normalizing Flows. The log-likelihood of the original cosmological field can be recovered by summing over the log-likelihood of each wavelet term. 
This decomposition allows us to separate the information from different scales and identify 
distribution shifts in the data such as unknown scale-dependent systematics. The resulting likelihood analysis can not only identify these types of systematics, but can also be made optimal, in the sense that the Multiscale Flow can learn the full likelihood at the field without any dimensionality reduction.
We apply Multiscale Flow to weak lensing mock datasets for cosmological inference, and show that it significantly outperforms traditional summary statistics such as power spectrum and peak counts, as well as novel Machine Learning based summary statistics such as scattering transform and convolutional neural networks. 
We further show that Multiscale Flow is able to identify distribution shifts not in the training data such as baryonic effects. Finally, we demonstrate that Multiscale Flow can be used to generate realistic samples of weak lensing data.
\end{abstract}
\vspace{0.35cm}

  \end{@twocolumnfalse} % Method A for two-column formatting
] % Method A for two-column formatting

%\begin{multicols}{2} % Method B for two-column formatting (doesn't play well with line numbers), comment out if using method A

%%%%%%%%%%%%%%%  Main text   %%%%%%%%%%%%%%%
% \linenumbers
\section{Introduction}
Numerous upcoming cosmological weak lensing
surveys such as \href{https://sci.esa.int/web/Euclid}{Euclid},  
%Dark Energy Spectroscopic Instrument \href{https://www.desi.lbl.gov/}{(DESI)}, 
the \href{https://lsst.org}{Vera Rubin Observatory} (Rubin), or  \href{https://roman.gsfc.nasa.gov/}{Nancy Grace Roman Space Telescope} (Roman) hold the promise of 
revolutionizing our understanding
of the universe, its origins, content, and its future evolution. 
How to efficiently extract the maximum amount of cosmological information from these data is a long-standing question in large-scale structure (LSS) analysis. Due to the high-order correlations induced by nonlinear gravitational effects, the late-time cosmological fields are highly non-Gaussian with no tractable likelihood functions. Extracting information from these non-Gaussian fields has been mainly attempted through a limited set of summary statistics, with the most popular ones being the N-point correlation functions \cite[e.g.,][]{peebles1980large,peebles1975statistical,sefusatti2006cosmology,semboloni2011weak,fu2014cfhtlens}. However, while the two-point function is a natural choice even 
in the nonlinear regime, higher-order correlation functions are significantly more difficult to use due to the large number of statistical coefficients, large variance and high sensitivity to outliers \cite{kendall1946advanced}. 
Numerous other statistics have been proposed, including correlation functions on transformed or marked fields \cite{neyrinck2009rejuvenating,white2016marked}, peak counts \cite{Jain2000statistics,kratochvil2010probing}, void statistics \cite{white1979hierarchy,pisani2019cosmic}, Minkowski functionals \cite{mecke1994robust}, scattering transform coefficients \cite{Cheng2020a,allys2020new}, statistics learned by convolutional neural networks (CNNs) and other NNs \cite{fluri2018cosmological,Charnock18,Makinen2021a,Gupta2018a,jeffrey2021likelihood}, and many others. 
%In each case, one must estimate their probability distribution as a function of cosmological and nuisance parameters, either by approximating it with a multi-variate Gaussian, or with Likelihood Free Inference \cite{Alsing2018a,Alsing2019a} or Simulation Based Inference \cite{Cranmer2020a} techniques. 
These analyses have the same underlying issues of summary statistics being ad-hoc and potentially sub-optimal. They 
require 
building effective likelihood functions from summary statistics
using multi-variate Gaussian or Simulation-Based Inference (SBI) methods \cite{Cranmer2020a}, which can be costly when the number of summaries is large. An alternative approach is using the reconstruction of initial conditions and estimating the field-level likelihood function by marginalizing over all possible initial conditions using a variety of methods such as sampling or optimization \cite{Jasche2013a, Kitaura2013a, Wang2014a,Seljak2017a,2022Porqueres}. 
These methods are expensive because they perform reconstructions or sampling of 
3-dimensional fields. They are also  not well matched to the problem when the data is 2-dimensional, such as weak lensing.  
%with $p(x|y) = \int p(x|y,z)p(z|y) dz$, where $x$ is the cosmological data, $y$ represent cosmological parameters, and $z$ denote linear density modes. 
%However, performing this marginal integral has proven to be difficult and expensive. To avoid this high-dimensional integral, one can use an unbiased estimator of the cosmological parameters based on the maximum a posteriori (MAP) estimation of the initial condition \cite{Seljak2017a, Modi2018a, bayer2022joint}, but it can still be expensive and sub-optimal to estimate the posterior.

Recently, Dai \& Seljak \cite{Dai2022a} proposed directly learning the field-level data likelihood with Normalizing Flows (NFs). This approach does not require compressing the data into a low-dimensional summary statistic, and instead tries to extract all the information in the data from the field-level likelihood. Unlike the 3-d reconstructions, this approach does not require evaluating the high dimensional integral, and computes the likelihood function in a single forward pass of the flow network. Unlike SBI, it uses 
field level likelihood instead of 
summary statistics, performing Simulation Based Likelihood Inference (SBLI). 
To reduce the degrees of freedom when modeling the high-dimensional likelihood of the data they enforce translation and rotation symmetry into the NF. The resulting Translation and Rotation Equivariant Normalizing Flow (TRENF) agrees well with the analytical solution on Gaussian Random Fields, and it leads to significant improvement over the standard power spectrum analysis on nonlinear matter fields from N-body simulations \cite{Dai2022a}. Similarly, NFs with different architectures have been applied to neutral Hydrogen (HI) maps for fast sample generation and cosmological inference \cite{Hassan2021,friedman2022higlow}. 
%Instead of learning the field-level likelihood functions, one can also directly estimate the 1D marginal posteriors from the fields with NNs \cite{villaescusa2021multifield,villanueva2022learning}, assuming that the marginal posterior is Gaussian.

Despite the differences in these LSS analysis methods, they all face the same challenge of robustness: how do we know which information is reliable, and which is not, if it is corrupted by effects that are ignored or inaccurately modeled? 
How do we detect distribution shifts in the actual data that were not in the training data? For example, most of these methods require accurate predictions from simulations, yet different hydrodynamical simulations and baryon models are not quite consistent with each other \cite{elahi2016nifty, huang2019modelling}. Villaescusa-Navarro et al. \cite{villaescusa2021multifield} train CNNs to predict cosmological parameters from gas temperature maps. They find that their model, trained using IllustrisTNG simulations \cite{pillepich2018simulating}, fails dramatically when applied to gas maps produced by SIMBA simulations \cite{dave2019simba}, due to the different subgrid models used in these two simulations. While marginalizing over the baryon parameters, subgrid models and various systematic effects are helpful and necessary, there is no guarantee that current baryon and systematic models span all potential realistic scenarios.  %Additional cares need to be taken to avoid introducing bias into the parameter inference.

%remove high peaks, scale cut. Apart from directly eliminating the data, scale comparison, Other methods include robust summary statistics (AP test, cross correlation),

%To mitigate the impact of such modeling uncertainties, the simplest approach is to eliminate data points that may be severely affected by this uncertainty, so that limitations in small-scale modelling do not bias the inferred cosmology [e.g. see Krause et al. (2017) for the determination of the redshift-dependent angular scale cuts for the DES-Y1 analysis or see Taylor, Bernardeau & Kitching (2018) for another method relating angular scale cuts to physical (k) space]. 

One way to mitigate the impact of such modeling uncertainties is by separation of scales, with very small-scale information likely being contaminated by many astrophysical nuisance effects and observation systematics, and large-scale information likely being more robust. This strategy is widely used in current cosmological survey analyses of power spectrum or correlation function, for example by directly removing the small-scale information with scale cuts \cite[e.g.,][]{Krause2017dark,taylor2018k,Krause2021dark}, or by performing consistency checks between different scales \cite{doux2021dark}. The ability 
to perform a scale-dependent analysis is viewed as a distinct 
advantage of power spectrum or 
correlation function analysis 
when compared to other statistics. 
%Apart from scale separation, other techniques for making robust cosmological constraints include clipping the high density peaks \cite{simpson2011clipping,simpson2013clipping}, and using robust summary statistics such as geometrical effect \cite{alcock1979evolution, ramanah2019cosmological} and cross correlation between different observables \cite[e.g.,][]{baldauf2010algorithm, Lavaux2021bayesian}.

In this paper, we apply the scale separation idea to the field-level likelihood modeling with NFs. Specifically, we use a set of scale-separated basis functions to represent the pixelized data, and decompose the data likelihood function into the contributions from different scales. Performing consistency checks between different scales enables us to decide what scale to include and what to exclude. 
%, performing robustness as a function of scale. 
While the Fourier basis is theoretically sound and widely used in such analysis, its kernels are not local in pixel space and require additional procedures in the presence of survey masks \cite{Dai2022a}. In this work we use wavelet basis, which is localized in both real space and Fourier space, allowing us to easily handle the survey mask and to separate the signals from different physical scales. Such decomposition is also known as Multiresolution Analysis (MRA) in image processing. Furthermore, 
our hierarchical analysis also 
combines likelihood information 
from different scales to achieve 
optimality in the limit of sufficient training data. 

%The plan of the paper is as following. We begin with a brief introduction to MRA and wavelet transforms in Section \ref{sec:MRA}. Then we develop normalizing flows to model the likelihood function of different wavelet components in Section \ref{sec:MF}. We name the model Multiscale Flow. In Section \ref{sec:WL} we apply Multiscale Flow to weak lensing cosmological analysis, both with and without baryon. We compare its performance with standard summary statistics power spectrum and peak count, as well as recently proposed novel statistics scattering transform and CNN. We present the cosmological constraints results in Section \ref{sec:result}. Finally we conclude in Section \ref{sec:conclusion}.

\section{Multiresolution Analysis with Fast Wavelet Transform}
\label{sec:MRA}

In this section we briefly introduce Multiresolution Analysis (MRA), which hierarchically decomposes the data into components at different scales, allowing us to separate the information from different scales and study them individually. This is particularly beneficial for cosmological analysis, since on large scales the universe can be modeled with simple physics and the data analysis is robust, while on small scales modeling the structure formation is harder due to nonlinear gravitational and astrophysical effects.

MRA is usually performed with Fast Wavelet Transform (FWT) \cite{mallat1989theory}. While similar in concept to the Fourier basis, wavelet bases are constructed to be localized spatially, which is beneficial when analyzing maps with irregular footprints. Wavelet transform has been widely used in astronomical image processing \cite{starck2007astronomical} and statistical description of cosmological fields \cite{Cheng2020a,allys2020new}. In this work, we focus on decimated wavelet transform, which preserves the dimensionality of the data and can be viewed as a special kind of NF transforms.

The basic idea of FWT is to recursively apply low-pass filters (also called scaling functions) and high-pass filters (also called wavelet functions) to the data. In each iteration, the data $x_{2^n}$ with resolution $2^n$ is decomposed into a low-resolution approximation $x_{2^{n-1}}$, and detail coefficients of the remaining signal $x^d_{2^{n-1}}$:
\begin{eqnarray}
\label{eq:FWT}
    x_{2^{n-1}} =& (\phi \ast x_{2^n}) \downarrow 2\\
    x^d_{2^{n-1}} =& (\psi \ast x_{2^n}) \downarrow 2
\end{eqnarray}
where $\phi$ is the low pass filter (scaling function), $\psi$ is the high pass filter (wavelet function), $\ast$ is the convolution operation, and $\downarrow 2$ is the operator to downsample the data by a factor of 2: $(x\downarrow 2)_{i,j} = x_{2i,2j}$. This is equivalent to a convolution with stride 2. For a 2D map $x_{2^n}$, we have three high pass filters to match the dimensionality, and the dimension of $x^d_{2^{n-1}}$ is $3\times 2^{n-1} \times 2^{n-1}$. Then the low-resolution data $x_{2^{n-1}}$ is passed to the next iteration and treated as the input for further decomposition. Note that this decomposition is bijective and in each iteration the input data can be reconstructed with the inverse wavelet transform.

In this work, we use Haar wavelet \cite{Haar1910theorie}, the simplest and the most spatially localized wavelet function. Its scaling function and wavelet function can be represented by the following $2\times 2$ kernel in real space:
\begin{alignat}{4}
\phi =& \frac{1}{4}
\begin{bmatrix}
1 & 1 \\
1 & 1 
\end{bmatrix} &&, \ 
&&\psi_1 = \frac{1}{2}
&&\begin{bmatrix}
1 & 1 \\
-1 & -1 
\end{bmatrix},\\
\psi_2 =& \frac{1}{2}
\begin{bmatrix}
1 & -1 \\
1 & -1 
\end{bmatrix} &&,\ 
&&\psi_3 =
&&\begin{bmatrix}
1 & -1 \\
-1 & 1 
\end{bmatrix},
\end{alignat}
where we have scaled the scaling function such that $x_{2^{n-1}}$ is exactly the low-resolution version of $x_{2^{n}}$ by taking the average of every $2\times 2$ patch. The localized kernel of the Haar wavelet allows us to handle the survey mask easily, but our method can be generalized to other more complicated wavelet transforms, e.g., Daubechies wavelets \cite{daubechies1988orthonormal}.

With MRA, the log-likelihood of a map $x_{2^n}$ with resolution $2^n$ can be rewritten into an auto-regressive form as
\begin{eqnarray}
\label{eq:decomposition}
    \log p(x_{2^n}|y) =& \log p(x_{2^{n-1}},x^d_{2^{n-1}}|y) \nonumber\\
    =& \log p(x_{2^{n-1}}|y) + \log p(x^d_{2^{n-1}}|x_{2^{n-1}},y) \nonumber\\
    =& \log p(x_{2^{n-2}}|y) + \log p(x^d_{2^{n-2}}|x_{2^{n-2}},y) \nonumber\\
    & + \log p(x^d_{2^{n-1}}|x_{2^{n-1}},y) \nonumber\\
    =& \cdots \nonumber\\
    =& \log p(x_{2^k}|y) + \sum_{m=k}^n \log p(x^d_{2^m}|x_{2^m},y) , 
\end{eqnarray}
where $2^k$ is the scale where we stops the decomposition, and $k$ can be any integer between $0$ and $n$. In practice, we can choose $k$ such that it corresponds to the scale that either has extracted 
all the information from the data, or is large enough not to be affected by unknown small-scale systematic effects.

\section{Multiscale Flow}

\subsection{Normalizing Flows}
%\textcolor{red}{Copied from TRENF paper}
Flow-based models provide a powerful framework for density estimation \cite{dinh2016density, papamakarios2017masked}
and sampling \cite{kingma2018glow}. These models map the data $x$ to latent variables $z$ through a sequence of invertible transformations $f = f_1 \circ f_2 \circ ... \circ f_n$, such that $z = f(x)$ and $z$ is mapped to a base distribution $\pi(z)$. The base distribution $\pi(z)$ is normally chosen to be a Gaussian
with zero mean and unit variance, $\pi(z)=\mathcal{N}(0,I)$. The probability density of data $x$ can be evaluated using the change of variables formula:
\begin{eqnarray}
    \label{eq:flow}
    p(x) =& \pi(f(x)) \left|\det \left(\frac{\partial f(x)}{\partial x}\right)\right| \nonumber \\
    =& \pi(f(x)) \prod_{l=1}^n \left|\det \left(\frac{\partial f_l(x)}{\partial x}\right)\right| .
\end{eqnarray}
To sample from $p(x)$, one first samples latent variable $z$ from $\pi(z)$, and then transform variable $z$ to $x$ through $x=f^{-1}(z)$. The transformation $f$ is usually parametrized with neural networks $f_{\phi}$, and the parameters $\phi$ are normally estimated using Maximum Likelihood Estimation (MLE):
\begin{equation}
\label{eq:MLE}
    \phi^* = \argmax_{\phi}\ \frac{1}{N}\sum_{i=1}^N\log p_{\phi}(x_i) ,
\end{equation}
where the data likelihood $p(x)$ is given by Equation \ref{eq:flow}. The MLE solution minimizes the Kullback-Leibler (KL) divergence between the model distribution $p_{\phi}(x)$ and the true data distribution. The parameterization of $f$ must satisfy the requirements that the Jacobian determinant $\det (\frac{\partial f_l(x)}{\partial x})$ is easy to compute for evaluating the density, and the transformation $f_l$ is easy to invert for efficient sampling.

In cosmological analysis we are interested in the likelihood function $p(x|y)$, which can be estimated using conditional Normalizing Flows (NFs). In conditional NFs the flow transformation is dependent on the conditional parameters $y$, i.e., $f=f_{\phi,y}$. We discuss below how we parametrize and train the conditional flow $f_{\phi,y}$. 

\subsection{Multiscale Flow}

\begin{figure*}[ht]
    \centering
    \includegraphics[width=\linewidth]{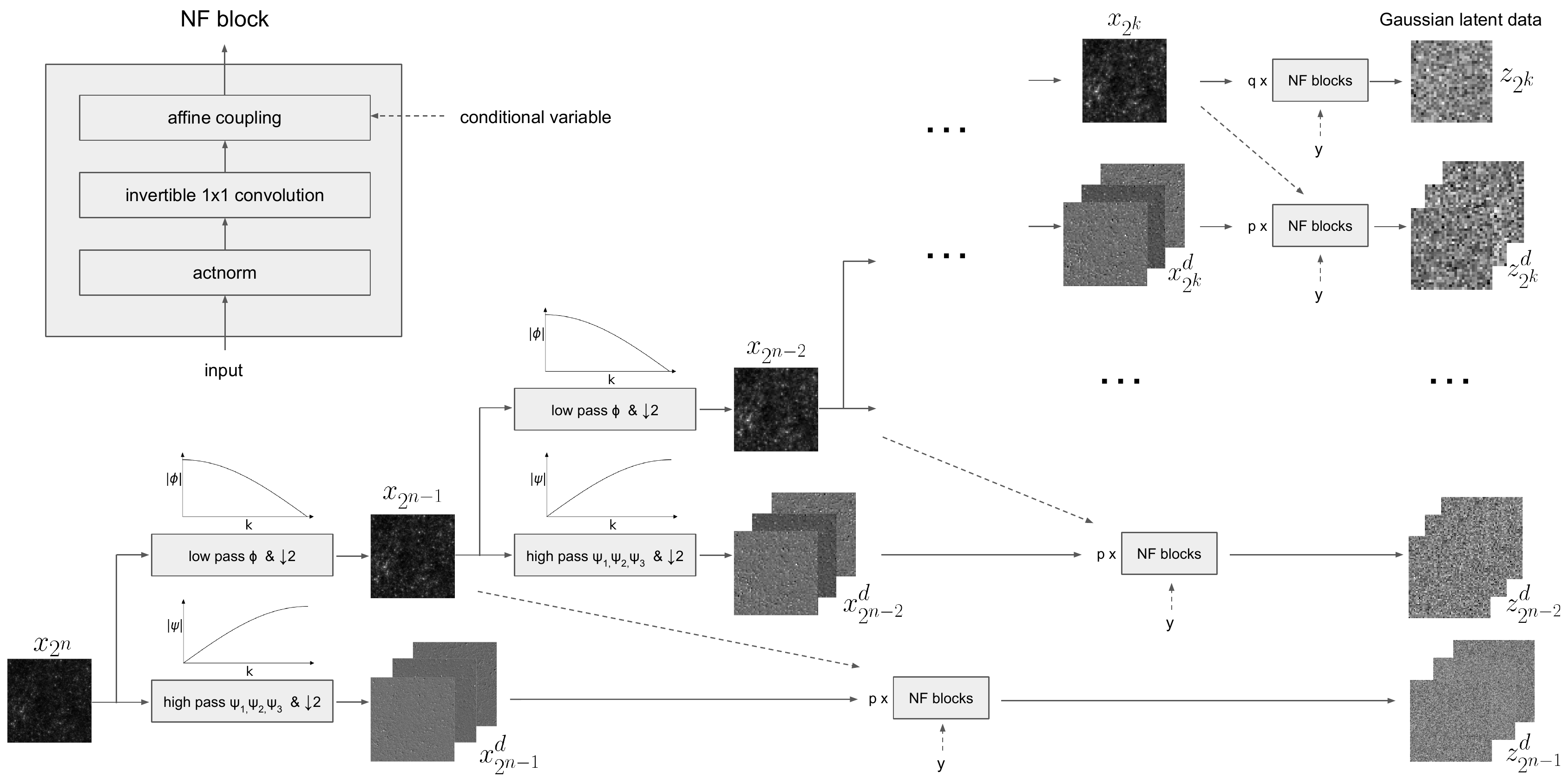}
    \caption{Illustration of Multiscale Flow model. The input map $x_{2^n}$ with resolution $2^n$ is iteratively processed with a set of low pass filters ($\phi$), high pass filters ($\psi_1$, $\psi_2$, $\psi_3$) and downsampling ($\downarrow 2$), resulting in a series of detailed maps $x_{2^{n-1}}^d, x_{2^{n-2}}^d, \cdots, x_{2^k}^d$ and an approximation map $x_{2^k}$. These maps are then transformed by several NF blocks to Gaussian latent maps $z_{2^{n-1}}^d, z_{2^{n-2}}^d, \cdots, z_{2^k}^d, z_{2^k}$, where each NF block is composed of an actnorm layer, an invertible $1\times 1$ convolution, and an affine coupling layer (Equation \ref{eq:affine1}, \ref{eq:affine2}), as shown on the top left of this figure. The NF transformation is conditioned on the conditional variable $y$ and approximation maps, which are represented by dashed arrows in the illustration. The log-likelihood of the input map $x_{2^n}$ can be calculated with Equation \ref{eq:decomposition}.}
    \label{fig:MF}
\end{figure*}

With the likelihood decomposition Equation \ref{eq:decomposition}, our task now is to build NFs to model different likelihood terms separately. 
%The large scale term $\log p(x_{2^k}|y)$ can be modeled with TRENF \cite{Dai2022a}, so in this section we will mainly focus on the conditional likelihood term $\log p(x_{2^m}^d|x_{2^m},y)$. 
For simplicity, we will drop the subscript $2^m$ in this section, and simply refer to the conditional likelihood term $\log p(x_{2^m}^d|x_{2^m},y)$ as $\log p(x^d|x,y)$. The model described here is similar to Wavelet Flow \cite{yu2020wavelet}, even though they are developed independently.
Following Glow \cite{kingma2018glow}, our flow transformation $f(x|y)$ consists of multiple block flows, where each block consists of an actnorm, an invertible $1\times 1$ convolution, and an affine coupling layer (Fig. \ref{fig:MF}). \\
\textbf{Actnorm}: The actnorm layer applies an affine transformation per channel, similar to batch normalization \cite{ioffe2015batch}, but its scale and bias parameters are initialized such that the output has zero mean and unit variance per channel given an initial minibatch of data, and then these parameters are treated as regular trainable parameters. \\
\textbf{Invertible $1\times 1$ convolution}: The invertible $1\times 1$ convolution is a learnable $C\times C$ matrix (where $C$ is the number of channels) that linearly mixes different channels. \\
\textbf{Affine coupling}: The affine coupling layer firstly splits the data $x^d$ to $x^{d1}$ and $x^{d2}$ based on the channels, and then applies pixel-wise affine transformation to $x^{d2}$, with scale and bias given by $x^{d1}$:
\begin{eqnarray}
\label{eq:affine1}
(\log s,\ t) =& \mathrm{CNN}(x^{d1}, x, y) \\
\label{eq:affine2}
z^{d2} =& \exp(\log s) \cdot x^{d2} + t ,
\end{eqnarray}
where $\log s$ and $t$ are scale and bias coefficient maps with the same dimensionality as $x^{d2}$, and CNN is a learned function parametrized by a convolutional neural network. The dependence of conditional parameter $y$ is modeled by introducing gating into CNN, i.e., each channel of CNN is scaled by a value between $0$ and $1$ which is determined by parameter $y$. This gating allows the conditional variable $y$ to determine the relative weights between different features (channels). 
The output of the affine coupling layer is the concatenation of $x^{d1}$ and $z^{d2}$. In other words, the affine coupling layer applies an affine transformation to $x^{d2}$ and leaves $x^{d1}$ unchanged. In this paper, we consider 2D maps, so at each scale $x^d$ contains $3$ maps (channels). We set the first channel to be $x^{d1}$, and the other two channels to be $x^{d2}$.

To summarize, a Multiscale Flow consists of multiple NFs, and each NF models one term of the likelihood decomposition (Equation \ref{eq:decomposition}) separately. The large-scale term $\log p(x_{2^k}|y)$ is modeled by $q$ flow blocks, and each other term $\log p(x_{2^m}^d|x_{2^m},y)$ is modeled with $p$ flow blocks, where $p$ and $q$ are hyperparameters in the model. Note that all of these NFs can be trained independently in parallel to speed up the training process.

\subsection{Training}
\label{subsec:training}

Following Dai \& Seljak \cite{Dai2022a}, we adopt a two-stage training strategy in this work: we first train the NF with the generative loss, which minimizes the negative log-likelihood and is the standard loss function of NF (Equation \ref{eq:MLE} with conditional variable y):
\begin{equation}
\label{eq:Lg}
\mathcal{L}_{\mathrm{g}} = -\frac{1}{N}\sum_{i=1}^N\log p(x_i|y_i) .
\end{equation}
The generative loss is suitable for sampling and density estimation, but may lead to a biased or overconfident posterior\cite{Dai2022a}. To solve this issue they propose further optimizing the posteriors by training the model with the discriminative loss,
\begin{align}
\label{eq:Ld}
    \mathcal{L}_{\mathrm{d}} 
 =&
 -\frac{1}{N}\sum_{i=1}^N\log p(y_i|x_i) \nonumber \\
    =& 
    -\frac{1}{N}\sum_{i=1}^N \left[ \log p(x_i|y_i)+ \log p(y_i) - \log p(x_i) \right] ,
\end{align}
where the evidence $p(x)$ is estimated using Importance Sampling (IS): $\log p(x) \approx \log \frac{1}{M}\sum_{y_j \sim q(y|x)}^M \frac{p(x|y_j)p(y_j)}{q(y_j|x)}$, and $q(y|x)$ is chosen to be a Gaussian distribution with learned mean and fixed covariance matrix. However, we find that IS becomes inefficient when the number of parameters $y$ gets large and when the posterior becomes non-Gaussian. In this work, we notice that 
\begin{align}
\label{eq:grad}
\nabla_{\phi}\mathcal{L}_{\mathrm{d}}=&
 -\frac{1}{N}\sum_{i=1}^N\left( \vphantom{\mathop{\mathbb{E}}_{y_j \sim p_{\phi}(y|x_i)}}\nabla_{\phi}\log p_{\phi}(x_i|y_i)-\right. \nonumber\\
  &\qquad\qquad\quad \left. \mathop{\mathbb{E}}_{y_j \sim p_{\phi}(y|x_i)}\nabla_{\phi}\log p_{\phi}(x_i|y_j)\right),
\end{align}
where we have used a trick that is commonly seen in the training of energy-based models. Its derivation can be found in \citep{song2021train}. In the training, we replace the expectation with a single Monte Carlo sample of the posterior $p(y|x_i)$, and we obtain these samples by running a Hamiltonian Monte Carlo (HMC) sampler \cite{duane1987hybrid}. These samples are saved, and then updated with a few HMC steps every epoch of training \cite{tieleman2008training}.
An advantage of this gradient formula compared to naively evaluating Eq \ref{eq:Ld}
%including the KL divergence $D_{\mathrm{KL}}(p(y|x)\parallel p(y))$ term 
is that instead of evaluating the evidence term $\log p(x) = \log \int p(x|y) p(y) dy$, we now evaluate $\int \log p(x|y) p(y|x) dy$. The estimation of the former usually comes with a large variance, while the latter can be estimated with only a few HMC samples.

\begin{figure}[t]
    \centering
    \includegraphics[width=\linewidth]{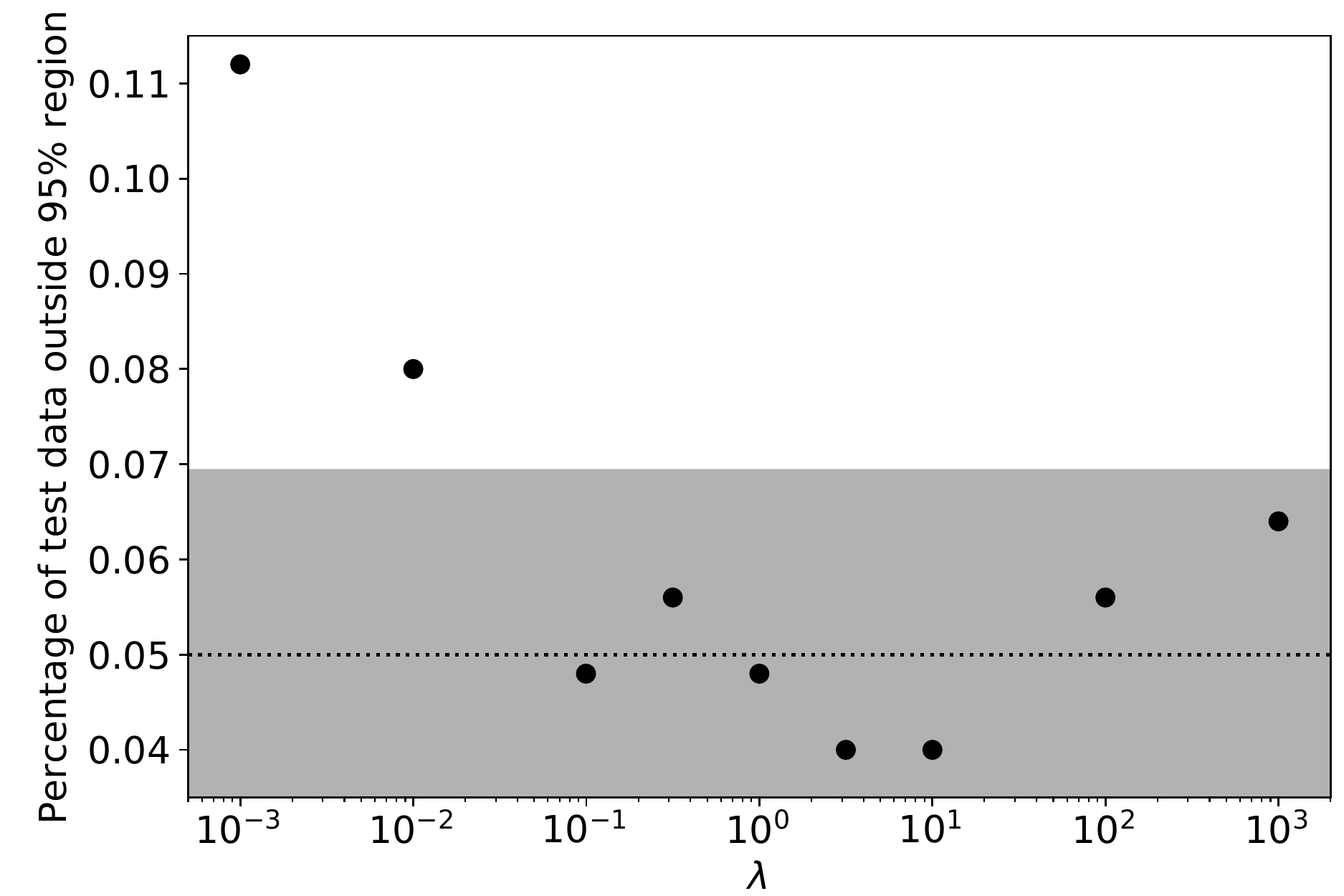}
    \caption{Percentage of test data that fall outside $95\%$ confidence region for different $\lambda$ values. A perfectly calibrated posterior has $5\%$ outliers. The shaded region shows the uncertainty due to finite number of test data. This measurement is made on weak lensing maps with $64^2$ resolution and $n_g=30 \mathrm{arcmin}^{-2}$ galaxy density.}
    \label{fig:lambda}
\end{figure}

After the generative training, we add this loss to the generative loss with a hyperparameter $\lambda$, 
\begin{equation}
    \mathcal{L}=\frac{1}{1+w\lambda}\mathcal{L}_{\mathrm{g}} +\frac{w\lambda}{1+w\lambda} \mathcal{L}_{\mathrm{d}},
\end{equation}
where $w=\frac{d_x}{d_y}$ is a prefactor to balance the dimension difference between the data and the parameter space, and we divide the loss by $1+w\lambda$ to normalize the weights. In Figure \ref{fig:lambda} we show the percentage of outliers in our posterior analysis with different $\lambda$ values. For very small $\lambda$ the posterior is too narrow (underestimated errors) and the loss is dominated by the first loss term (generative loss). For $\lambda > 0.1$ 
the posterior is well calibrated due to the second term $\mathcal{\tilde{L}}_{\mathrm{d}}$. In this paper, we use $\lambda=1$ to calibrate the posterior. %, which leads to a slightly conservative calibration (Table \ref{tab:coverage} and Figure \ref{fig:lambda}).

\section{Results}
\label{sec:result}

\subsection{Cosmological constraints from noisy weak lensing maps}

\begin{figure}[t]
    \centering
    \includegraphics[width=\linewidth]{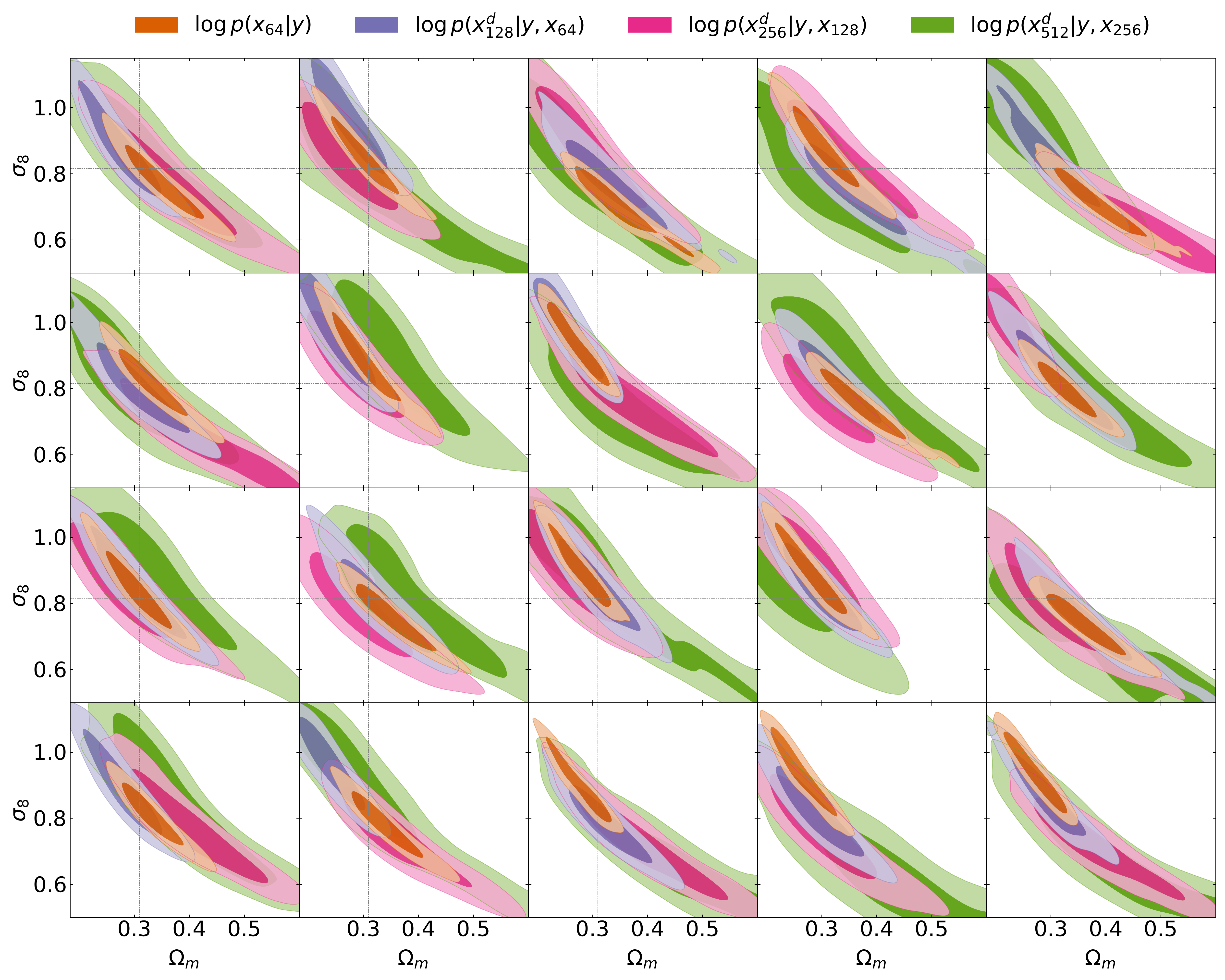}
    \caption{Multiscale Flow posterior comparison of different scales on 20 test data with galaxy number density $n_g=30 \mathrm{arcmin}^{-2}$.}
    \label{fig:posterior}
\end{figure}

\begin{table*}
  \caption{Comparison of the constraining power between different methods. The figure of merit is measured by the reciprocal of the $1\sigma$ confidence area on the $(\Omega_m, \sigma_8)$ plane, using a $3.5\times 3.5\  \mathrm{deg}^2$ convergence map.}
  \label{tab:FoM}
  \centering
  \begin{tabular}{>{\centering}c|>{\centering}c>{\centering}c>{\centering\arraybackslash}c}
    \toprule
    Method & $n_g=10 \mathrm{arcmin}^{-2}$ & $n_g=30 \mathrm{arcmin}^{-2}$ & $n_g=100 \mathrm{arcmin}^{-2}$\\% & noiseless\\ 
    \midrule
   Multiscale Flow $p(x_{512}|y)$ & \bf{89} & \bf{248} & \bf{740}\\% & \\
    Multiscale Flow $p(x_{256}|y)$ & 82 & 226 & 631\\% & \\
    Multiscale Flow $p(x_{128}|y)$ & 76 & 191 & 472\\% & \\
    Multiscale Flow $p(x_{64}|y)$ & 62 & 130 & 298\\% & \\
    power spectrum & 30 (30) & 52 (51) & 81 (79)\\% & 246 (107)\\
    %power spectrum (CNN paper) & (29) & (42) & (63) & - \\
    %power spectrum (ST paper, Fisher) & (20) & (40) & (67) & 253 (104)\\
    peak count & (40) & (85) & (137)\\% & 667(170)\\
    CNN & (44) & (121) & (292)\\% & (1201)\\
    scattering transform $s_0+s_1+s_2$ & ($\lesssim 50$) & ($\lesssim 140$) & ($\lesssim 329$)\\% & 3367 (1053)\\
    \bottomrule
  \end{tabular}
  \\
  \addtabletext{
  \begin{enumerate}
      \item Unless specified with Multiscale Flow, the analysis of other approaches are performed on maps with resolution $512^2$. 
      \item The numbers in parenthesis are estimated using maps with $1\ \mathrm{arcmin}$ Gaussian smoothing. We expect this smoothing to have little effect on constraining power estimation, because the small-scale modes are dominated by shape noise. This is also explicitly verified in the case of power spectrum, where we show FoM with and without smoothing. We have also verified that CNN produces comparable results with and without smoothing.
      \item The FoM of the scattering transform is estimated using the Fisher matrix, which is an upper limit of the true FoM according to the Cram\'er-Rao inequality. It has been shown that Fisher forecast could potentially overestimate the 1D parameter constraints by a factor of 2, due to the non-Gaussian distribution of the statistics. \cite{park2022quantification}.
  \end{enumerate}
  }
\end{table*}

We apply Multiscale Flow to $3.5 \times 3.5 \mathrm{deg}^2$ mock weak lensing convergence maps \cite{Ribli2019a} for field-level inference. We decompose the $512^2$ resolution map to four scales, with likelihood decomposition
\begin{eqnarray}
\label{eq:logL_WL}
    \log p(x_{512}|y) = &\log p(x_{64}|y) + \log p(x_{64}^d|x_{64},y) + \nonumber\\ 
    &\log p(x_{128}^d|x_{128},y) + \log p(x_{256}^d|x_{256},y) .
\end{eqnarray}
The posterior comparison of different scales on $20$ test maps with galaxy number density $n_g = 30 \mathrm{arcmin}^{-2}$ is shown in Figure \ref{fig:posterior}. The posterior constraints of all scales are consistent with the true cosmological parameters, which are shown as black lines. The constraining power of Multiscale Flow of different galaxy shape noise levels is shown in Table \ref{tab:FoM}. We list the figure of merit (defined as the reciprocal of the $1\sigma$ confidence area on the $(\Omega_m, \sigma_8)$ plane) of maps with different resolutions, and compare them with summary statistics power spectrum, peak count, scattering transform \cite{Cheng2020a}, and statistics learned by CNNs \cite{Ribli2019a}. Multiscale Flow achieves the best performance among all methods, outperforming power spectrum by factors of 3, 5 and 9 on galaxy densities $n_g=10, 30, 100 \mathrm{arcmin}^{-2}$, respectively. Multiscale Flow also achieves two to three higher constraining power when compared to peak counts, CNN, and scattering transform.

\subsection{Impact of baryons}

\begin{figure}[t]
    \centering
    \includegraphics[width=\linewidth]{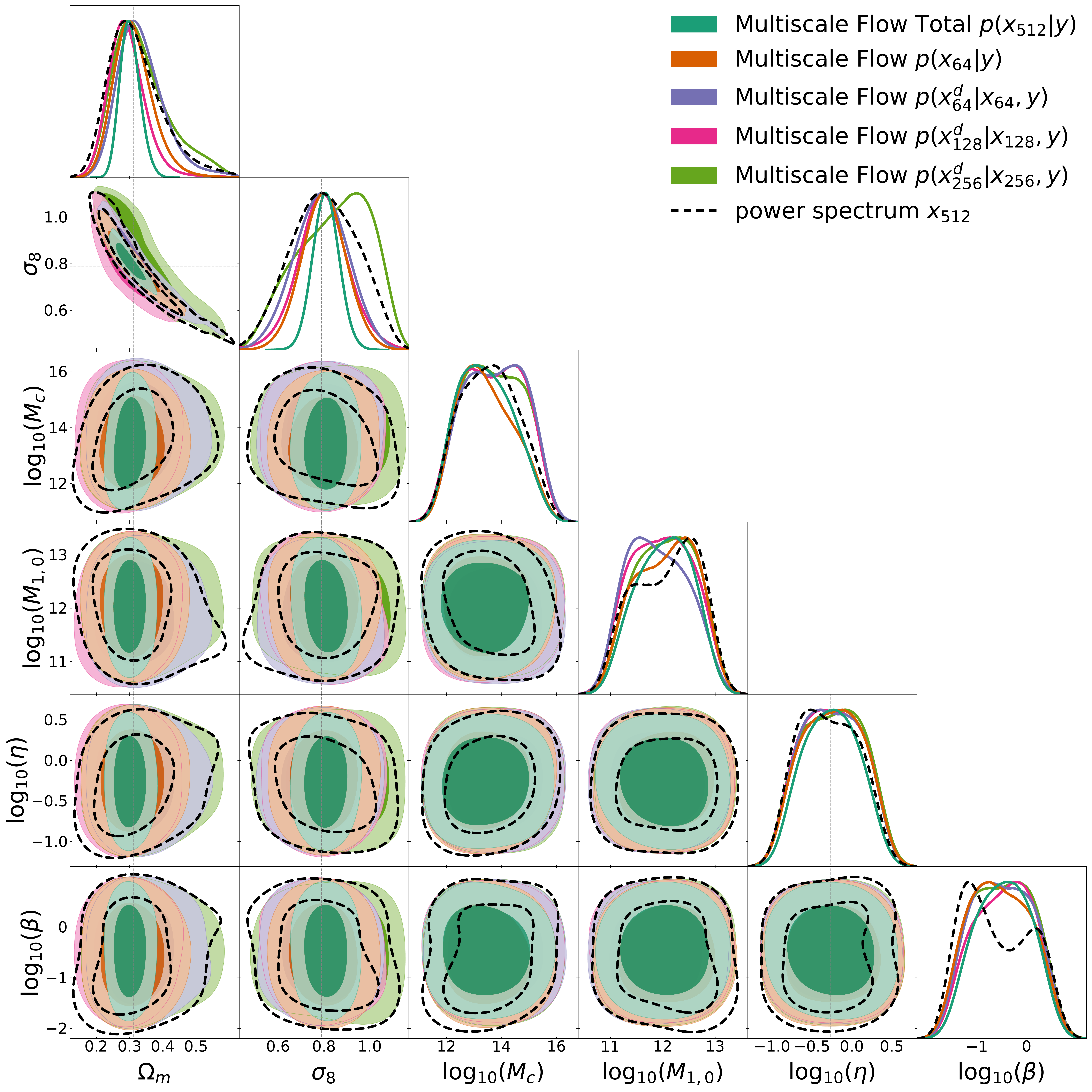}
    \caption{Comparison of posterior distributions between different scales of Multiscale Flow and power spectrum on a $3.5 \times 3.5 \mathrm{deg}^2$ convergence map with $n_g=20 \mathrm{arcmin}^{-2}$.}
    \label{fig:posterior_baryon}
\end{figure}

\begin{table*}[t]
  \caption{Similar to Table \ref{tab:FoM}, with baryonic effects. %\textcolor{red}{add peak count?}
  }
  \label{tab:FoM_baryon}
  \centering
  \begin{tabular}{>{\centering}c|>{\centering}c|>{\centering}c>{\centering}c>{\centering}c>{\centering\arraybackslash}c}
    \toprule
    & Method & $n_g=10 \mathrm{arcmin}^{-2}$ & $n_g=20 \mathrm{arcmin}^{-2}$ & $n_g=50 \mathrm{arcmin}^{-2}$ & $n_g=100 \mathrm{arcmin}^{-2}$\\ 
    \midrule\midrule
    \multirow{6}{*}{\shortstack{Fix baryon\\parameters at\\fiducial values}} & Multiscale Flow $p(x_{512}|y)$ & \bf{104} & \bf{203} & \bf{469} & \bf{787} \\
    & Multiscale Flow $p(x_{256}|y)$ & 99 & 186 & 408 & 654 \\
    & Multiscale Flow $p(x_{128}|y)$ & 87 & 155 & 319 & 471 \\
    & Multiscale Flow $p(x_{64}|y)$ & 68 & 112 & 210 & 306\\
    %&\rule{0pt}{3ex} power spectrum & - & (44) & (73) & (97) \\
    &\rule{0pt}{3ex} power spectrum & 41(41) & 61 (58) & 95(87) & 127(111) \\
    & CNN\tnote{1} & - & ($\sim 93$) & ($\sim 146$) & ($\sim 194$) \\
    %& CNN + power spectrum\tnote{1} & - & (102) & (177) & (240) \\
    \midrule
    \multirow{6}{*}{\shortstack{Marginalize\\over baryon\\parameters}} & Multiscale Flow $p(x_{512}|y)$ & \bf{84} & \bf{144} & \bf{254} & \bf{359}\\
    & Multiscale Flow $p(x_{256}|y)$ & 82 & 137 & 242 & 338 \\
    & Multiscale Flow $p(x_{128}|y)$ & 71 & 118 & 206 & 290 \\
    & Multiscale Flow $p(x_{64}|y)$ & 59 & 91 & 154 & 210 \\
    &\rule{0pt}{3ex} power spectrum & 34(33) & 48(48) & 68(65) & 84 (78) \\
    %&  power spectrum  & - & (12) & (15) & (19) \\
    & CNN\tnote{2} & - & ($\sim 77$) & ($\sim 109$) & ($\sim 136$) \\
    %& CNN + power spectrum\tnote{2} & - & ($\sim 84$) & ($\sim 132$) & ($\sim 169$) \\
    \bottomrule
  \end{tabular}
  \\
  \addtabletext{
  \begin{enumerate}
      \item When fixing the baryon parameters at fiducial values, the FoM of CNN are estimated from Lu et al. \cite{Lu2022simultaneously}. Lu et al. \cite{Lu2022simultaneously} estimated the $1\sigma$ area of a $1500 \mathrm{deg}^2$ survey, and we scale their results by the area ratio for a direct comparison with our experiments.
      \item For marginalizing over baryon parameters, simply rescaling the results of Lu et al. \cite{Lu2022simultaneously} by the area ratio underestimates the constraining power of CNN, due to the prior bounds of baryon parameters. Instead, we estimate its FoM by $\mathrm{\frac{FoM_{PS, marginal}}{FoM_{PS, fiducial}} FoM_{CNN, fiducial}}$. %Note that $\mathrm{\frac{FoM_{PS, marginal}}{FoM_{PS, fiducial}}} > \mathrm{\frac{FoM_{CNN, marginal}}{FoM_{CNN, fiducial}}}$ \citep{Lu2022simultaneously}, so this estimation may overestimate its FoM. 
  \end{enumerate}
  }
\end{table*}

\begin{table*}[t]
  \caption{Empirical coverage probability of posteriors from different methods, after marginalizing over baryon parameters. We report the percentage of test data that falls within the $68\%$ confidence region and the $95\%$ confidence region. A perfectly calibrated posterior should have $68\%$ and $95\%$ test data that fall in these two regions, respectively. %We observe overly conservative contours, suggesting the true Figure of Merit may be higher than quoted.
  }
  \label{tab:coverage}
  \vskip 0.15in
  \centering
  \begin{threeparttable}
  \begin{tabular}{>{\centering}c|>{\centering}c>{\centering}c>{\centering}c>{\centering\arraybackslash}c}
    \toprule
    Method & $n_g=10 \mathrm{arcmin}^{-2}$ & $n_g=20 \mathrm{arcmin}^{-2}$ & $n_g=50 \mathrm{arcmin}^{-2}$ & $n_g=100 \mathrm{arcmin}^{-2}$\\ 
    \midrule\midrule
    Multiscale Flow $p(x_{512}|y)$ & $67.2\%,\ 93.0\%$ & $72.7\%,\ 95.3\%$ & $71.1\%,\ 96.9\%$ & $71.1\%,\ 98.4\%$ \\% \bf{76} & \bf{143} & \bf{230} & \bf{383} \\
    Multiscale Flow $p(x_{256}|y)$ & $65.6\%,\ 94.5\%$ & $71.1\%,\ 95.3\%$ & $74.2\%,\ 96.1\%$ & $73.4\%,\ 97.7\%$\\
    Multiscale Flow $p(x_{128}|y)$ & $68.0\%,\ 94.5\%$ & $71.9\%,\ 96.1\%$ & $75.0\%,\ 95.3\%$ & $77.3\%,\ 97.7\%$\\
    Multiscale Flow $p(x_{64}|y)$ & $66.4\%,\ 96.9\%$ & $67.2\%,\ 96.1\%$ & $72.7\%,\ 96.1\%$ & $73.4\%,\ 97.6\%$ \\
    %\rule{0pt}{3ex} power spectrum &&&& \\
    \bottomrule
  \end{tabular}
  \end{threeparttable}
  \vskip -0.1in
\end{table*}

Next, we apply Multiscale Flow to mock weak lensing maps with baryonic physics included \cite{Lu2022simultaneously}. Similar to the previous experiment, these maps also have a resolution of $512^2$, and we adopt the same likelihood decomposition as Equation \ref{eq:logL_WL}. We have 6 physical parameters in total, i.e., cosmological parameters $\Omega_m$ and $\sigma_8$, and 4 baryon parameters \cite{Aric02020a}. The posterior distributions of Multiscale Flow and power spectrum of a test data with $n_g=20 \mathrm{arcmin}^{-2}$ are shown in Figure \ref{fig:posterior_baryon}. In Table \ref{tab:FoM_baryon} we compare the constraining power of Multiscale Flow, power spectrum, and CNN \cite{Lu2022simultaneously} on $(\Omega_m, \sigma_8)$ plane. With the presence of baryon physics, Multiscale Flow has $2.5 - 4$ times higher constraining power on cosmological parameters when compared to the power spectrum. It also outperforms CNN by a factor of 2. 

Unfortunately, due to the small area of the lensing map, all these methods cannot constrain baryon parameters very well (see also Figure 5 of Lu et al. \cite{Lu2022simultaneously} for CNN constraints), and the posterior is dominated by the prior bounds, especially in the cases of high shape noise. Therefore, marginalizing the baryon parameters has a small impact on the Figure of Merit. With smaller shape noise and a more powerful model, the posterior becomes more dominated by likelihood rather than the prior, and the degradation of FoM when marginalizing over the baryon parameters gets larger. This explains why the degradation of baryon marginalization is larger for Multiscale Flow compared to the power spectrum, and why the degradation is larger in high galaxy number density cases. 
However, it is important to 
recognize that with better 
statistical power, and simpler
baryonic models, we expect field level 
inference to be able to 
break the degeneracies between 
baryonic and cosmological parameters.

We apply Multiscale Flow to test data with fiducial parameters, and in Table \ref{tab:coverage} we report the percentage of test data with true cosmological parameters to fall in $68\%$ and $95\%$ confidence regions. In most cases the percentages are larger than the $68\%$ and $95\%$ expectation, suggesting that our posterior constraint is conservative.

\subsection{Identifying distribution shifts}

\begin{figure}
     \centering
     \begin{subfigure}[]{0.49\linewidth}
         \centering \includegraphics[width=\linewidth]{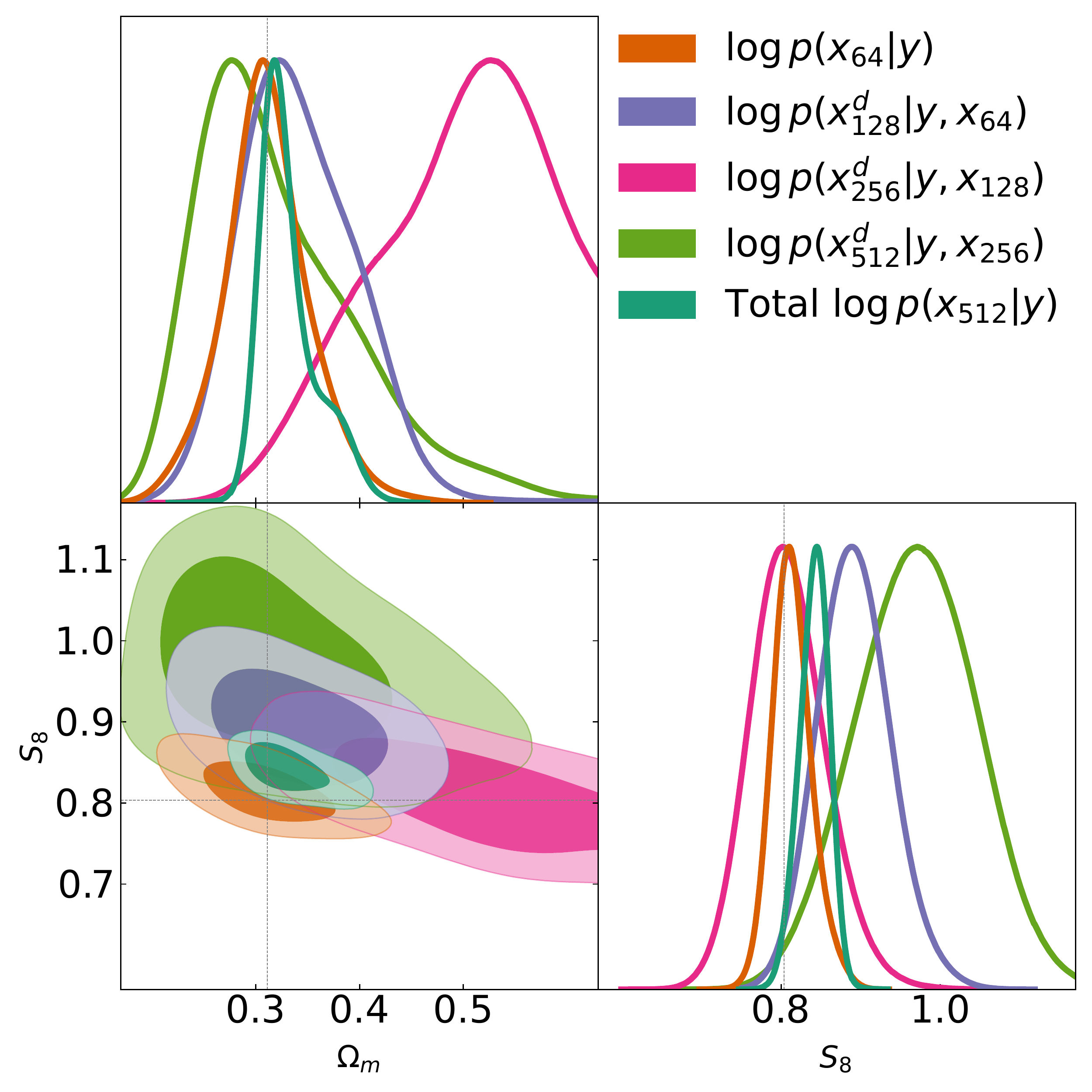}
     \end{subfigure}
     \hfill
     \begin{subfigure}[]{0.49\linewidth}
         \centering \includegraphics[width=\linewidth]{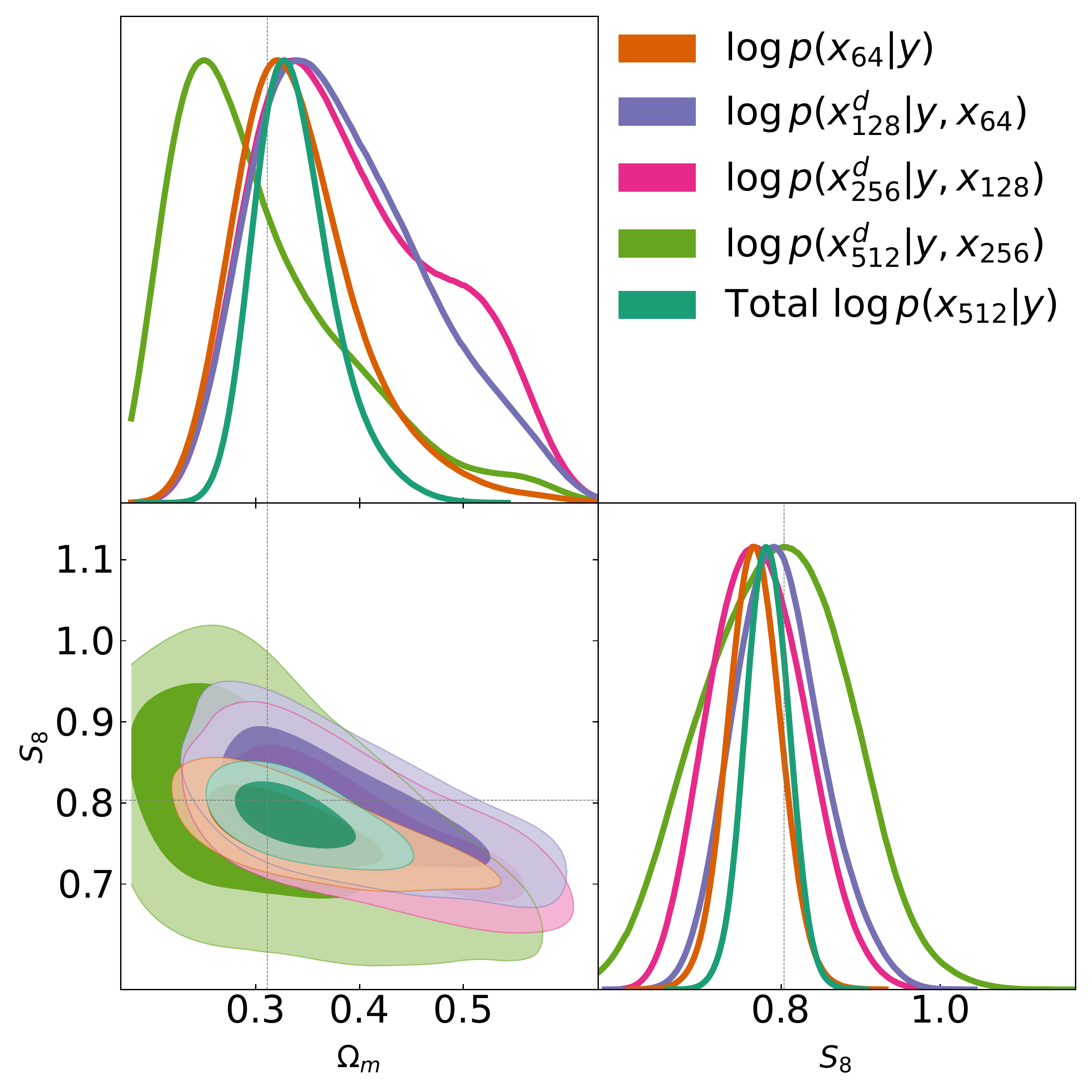}
     \end{subfigure}
     \hfill
     \begin{subfigure}[]{0.49\linewidth}
         \centering \includegraphics[width=\linewidth]{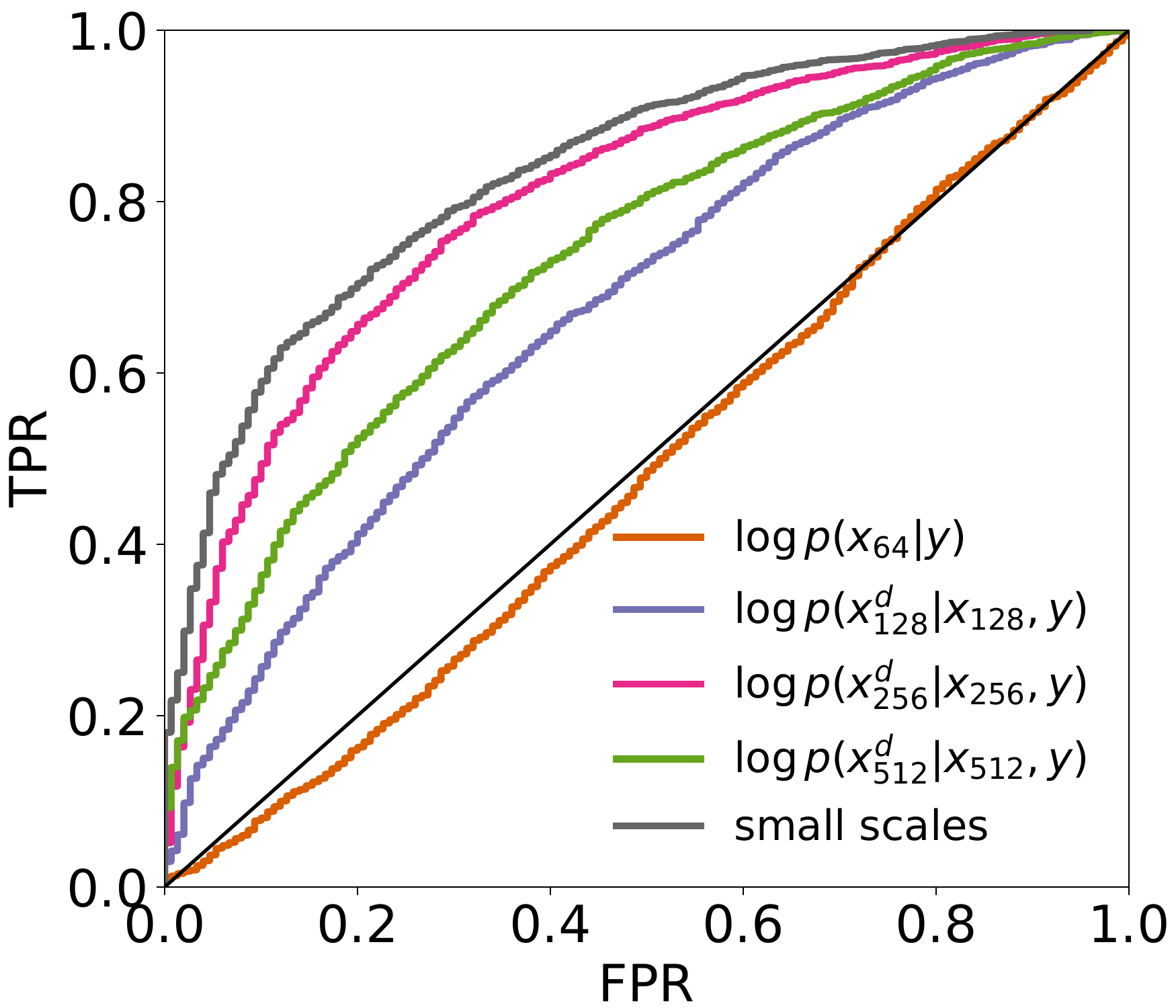}
     \end{subfigure}
     \hfill
     \begin{subfigure}[]{0.49\linewidth}
         \centering \includegraphics[width=\linewidth]{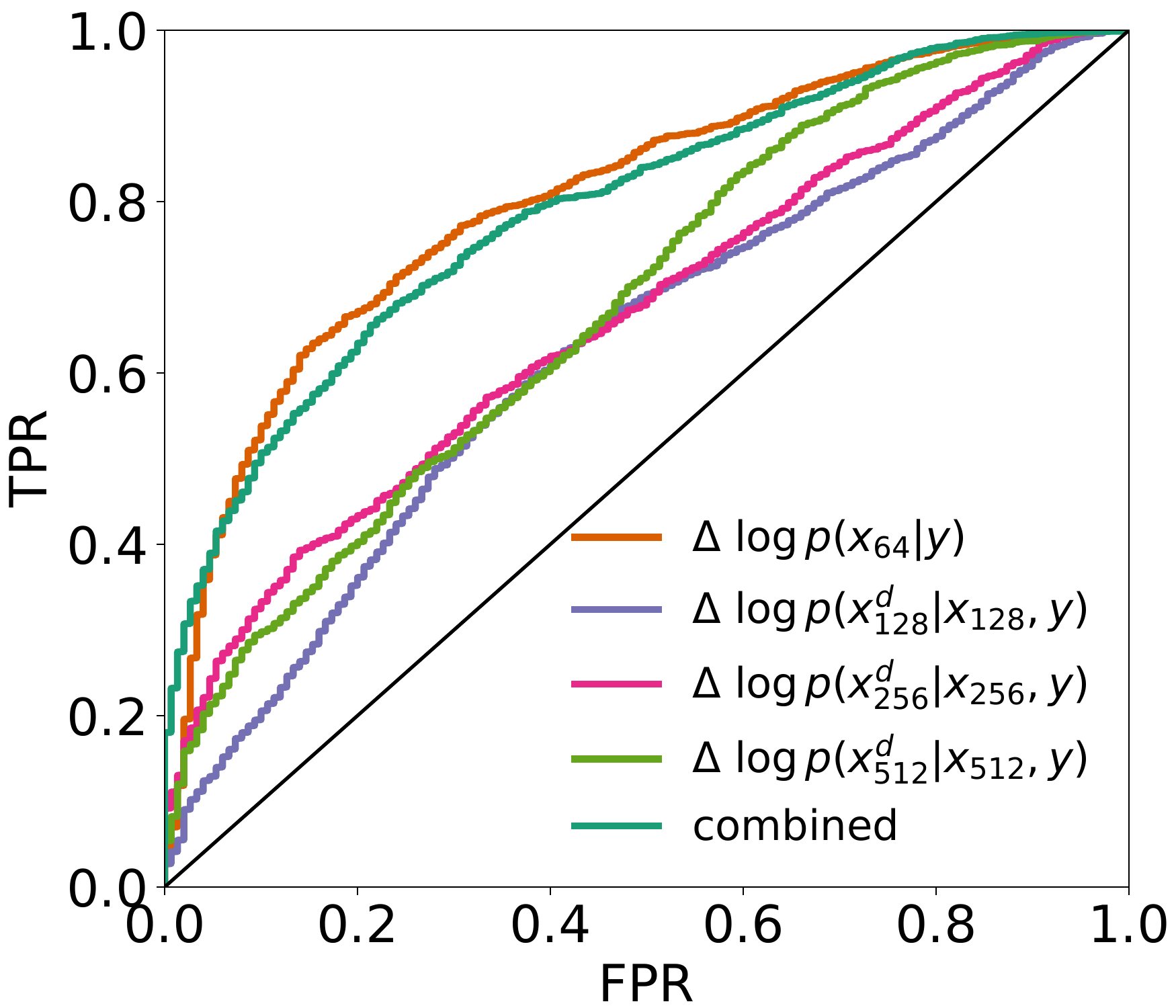}
     \end{subfigure}
     %\hfill
     %\begin{subfigure}[]{0.32\linewidth}
     %    \centering \includegraphics[width=\linewidth]{distribution_shift_PS.pdf}
     %\end{subfigure}
    \caption{Top panel: scale-dependent posterior analysis of a baryon-corrected convergence map using Multiscale Flow trained on dark-matter-only maps (left), and Multiscale Flow trained on BCM maps (right). Bottom panel: ROC curve of identifying distribution shift with $\log p$ (left) and $\Delta \log p$ (right). The "small scales" in the lower left panel represent combining the three small scale terms. In these experiments, we consider $30 \mathrm{arcmin}^{-2}$ galaxy shape noise.}
    \label{fig:distribution_shift}
\end{figure}

Identifying distribution shifts from unknown effects that are present in the data, but not in the training simulations, is one of the great challenges of modern Machine Learning. 
Here we propose two different methods to identify such shifts. In the first approach, we evaluate the likelihood value of test data at MAP $\log p(x|y_{\mathrm{MAP}})$ and compare it with the distribution of training data. If it is smaller than the typical likelihood values of training data, it is likely not in the typical set of training distribution. In the second approach, we use consistency of information as a function of scale to identify such shifts. Specifically, we evaluate 
\begin{equation}
\Delta \log p(x_m|y) = \log p(x_m|y_{\mathrm{MAP}}) - \log p(x_m|y_{\mathrm{MAP},m}),
\end{equation}
where $x_m$ is the data of a specific scale, $y_{\mathrm{MAP},m} = \argmax_y \log p(y|x_m)$ is the MAP of this scale, and $y_{\mathrm{MAP}}$ is the MAP of all the scales. If there are scale-dependent systematic effects that bias the posterior in different ways, we expect $y_{\mathrm{MAP}}$ and $y_{\mathrm{MAP},m}$ to be quite different, and $\Delta \log p(x_m|y)$ should be smaller compared to those of training data. 

As a simple example, we train the Multiscale Flow with dark-matter-only convergence maps \cite{Ribli2019a}, and apply the model to convergence maps with baryon physics included \cite{Lu2022simultaneously}. We show the posterior distributions from different scales in the upper left panel of Figure \ref{fig:distribution_shift}. The baryon physics modifies the matter distribution on small scales and biases the posterior constraints from small scales. In this case, naively combining all of the scales leads to a posterior constraint that is $2\sigma$ biased (dark green contour). The inconsistency of posterior between different scales suggests a presence of unknown systematics (baryon physics) that is not modeled in the training data. If we remove the small-scale information (because we believe the large scales are less likely to be affected by systematics), we can recover an unbiased constraint of cosmological parameters (orange contour).  
As a comparison, in the upper right panel, we show the posteriors from Multiscale Flow trained using maps with baryon physics. There is no distribution shift in this case and the information from the different scales is consistent.

In the bottom panel of Figure \ref{fig:distribution_shift}, we show the ROC curve of identifying this distribution shift with $\log p$ and $\Delta \log p$. As expected, the likelihood of large-scale term $\log p(x_{64}|y)$ cannot tell the difference between with and without baryon physics, while the likelihood of small-scale terms can be used for detecting the shifts. By combining all the small-scale terms, we get the best performance with AUROC of $0.84$. We also find that $\Delta \log p$ work equally well in this task. In this case the large-scale term $\Delta \log p(x_{64}|y)$ achieves the best performance with AUROC of $0.80$, because the small-scale constraints bias $y_{\mathrm{MAP}}$ away from $y_{\mathrm{MAP,64}}$. The two 
methods are essentially independent, and combining them further improves OoD detection. These maps have a small area ($3.5\times 3.5 \mathrm{deg}^2$), and the 2048 test data used in this experiment span a wide range of baryon parameters, of which many are likely indistinguishable from the no baryons given the sampling variance between the maps. We expect our OoD detection methods will work even better for sky surveys with larger areas and for models where baryonic effects are more significant.

\subsection{Sample generation and super-resolution}

\begin{figure*}[t]
     \centering
    \includegraphics[width=\linewidth]{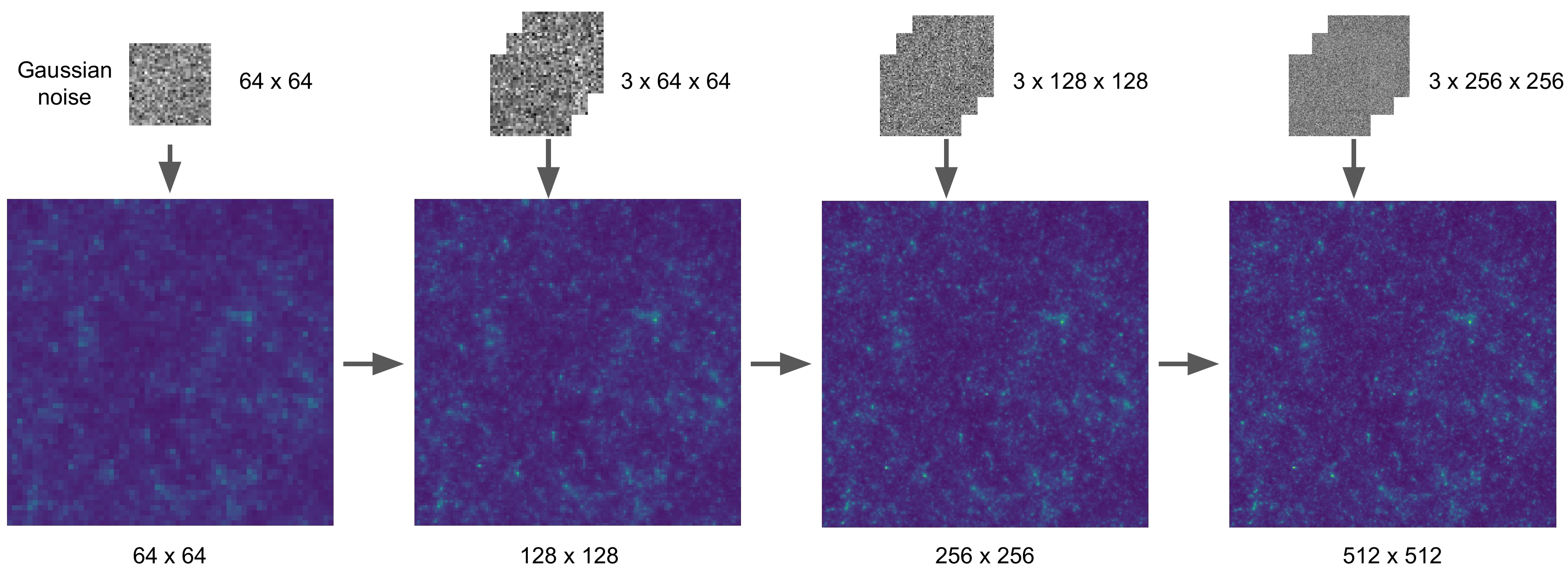}
    \caption{Illustration of Multiscale Flow sample generation (the reverse of Figure \ref{fig:MF}). The sample of the lowest resolution is first generated, and then small-scale information is gradually added. This process can also be viewed as super-resolution.}
    \label{fig:MF_sample}
\end{figure*}

\begin{figure}
     \centering
     \begin{subfigure}[]{0.48\linewidth}
         \centering \includegraphics[width=\linewidth]{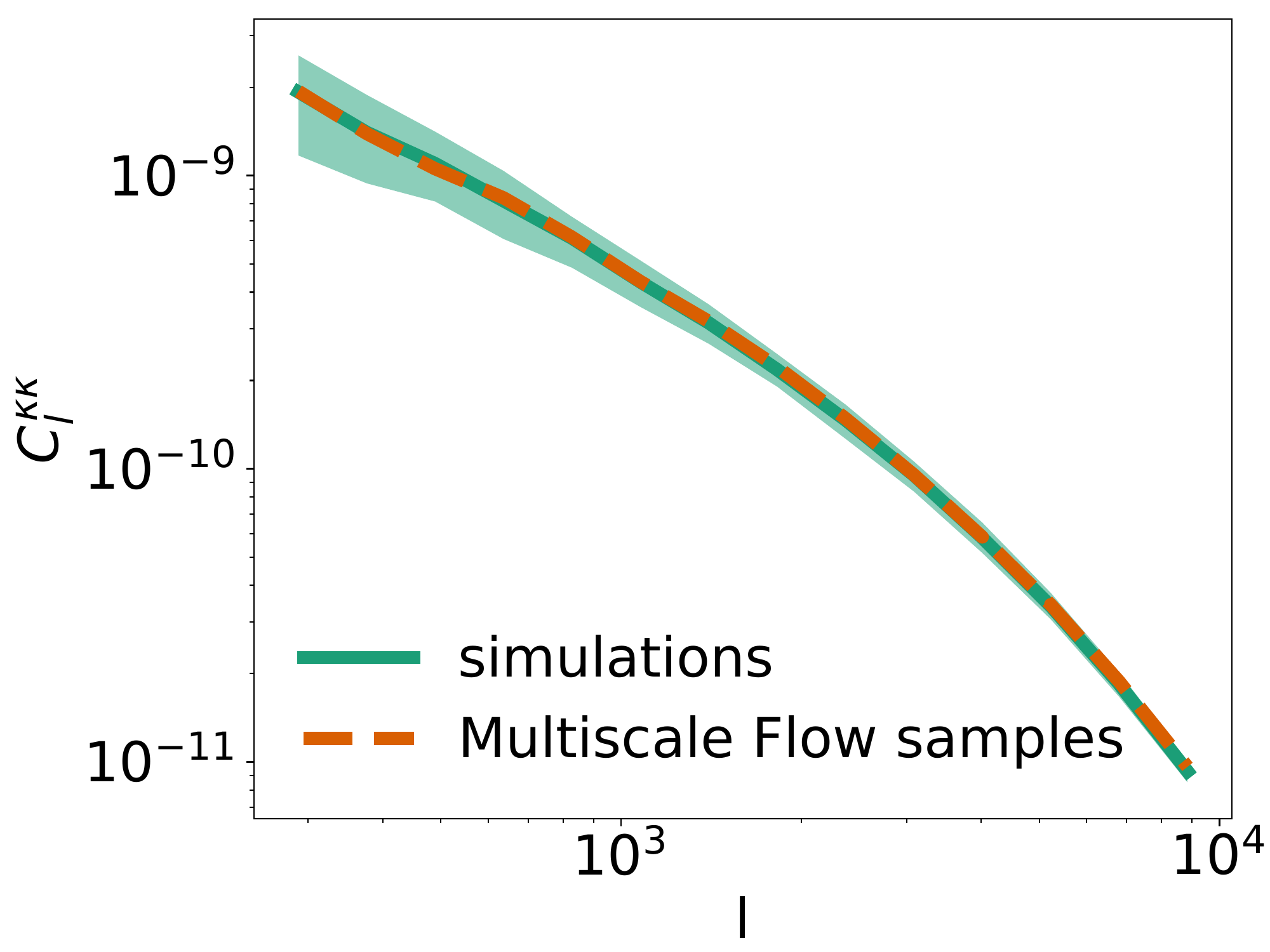}
     \end{subfigure}
     \hfill
     \begin{subfigure}[]{0.48\linewidth}
         \centering \includegraphics[width=\linewidth]{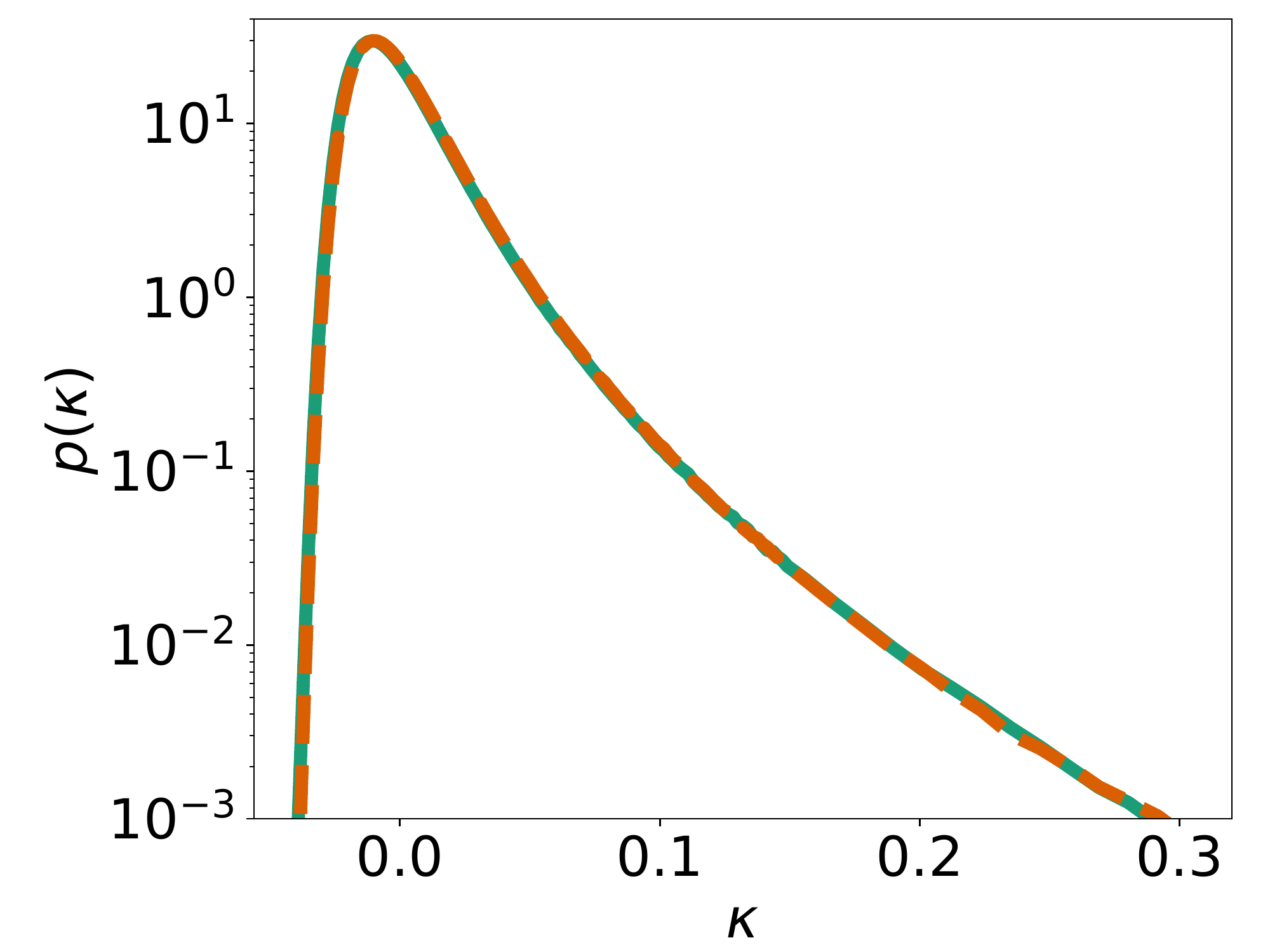}
     \end{subfigure}
    \caption{Comparison of the power spectrum (left) and pixel probability distribution function (right) between simulations and Multiscale Flow samples at fiducial cosmology.}
    \label{fig:samples_statistics}
\end{figure}

We show an example of sample generation with Multiscale Flow in Figure \ref{fig:MF_sample}. The process can also be viewed as iterative super-resolution of the low-resolution samples. In Figure \ref{fig:samples_statistics} we show that Multiscale Flow samples and test data agree well in terms of the power spectrum and pixel probability distribution function.
This demonstrates that Multiscale Flow samples can be used in lieu of expensive N-body simulations and ray tracing as a fast generator of mock data. 

\section*{Comparison with other machine learning models}

\subsection*{Comparison with discriminative models}

So far there are lots of works using machine learning models to extract cosmological information at the field level. Most of these works either train models to directly learn the posterior constraints \cite{villaescusa2021multifield,villanueva2022learning}, or build models to perform data compression $x \rightarrow s$ for cosmological inference, where the summary statistics $s$ can be a point estimate of the cosmological parameter \cite{fluri2018cosmological,fluri2019cosmological,Gupta2018a,Ribli2019a,Lu2022simultaneously,lu2023cosmological}, or simply a data vector that contains rich information about cosmological parameter $y$ \cite{Makinen2021a,fluri2022full}. These models are generally referred to as discriminative models.

Generative models, on the other hand, learn the data likelihood function $p(x|y)$, and then calculate the posterior distribution using Bayes rule. It has been suggested that while discriminative models have less asymptotic error, generative models have less sample complexity \cite{ng2002discriminative,Yogatama2017generative,zheng2023revisiting}. In other words, there can be two distinct regimes of performance as the training set size is increased. When the training size is small, the generative model achieves its asymptotic error much more quickly as data increases and can outperform the discriminative model, because the latter is more likely to overfit and requires more training data to converge. 

For the weak lensing dataset considered in this work, the training set size is relatively small ($2.9\times 10^4$ for maps without baryonic physics, and $7.7\times 10^4$ for maps with baryonic physics) compared to the dimensionality $d=512^2\approx 2.6\times 10^5$. This explains why Multiscale Flow, which learns the data likelihood function, outperforms CNN in Table \ref{tab:FoM} and \ref{tab:FoM_baryon}. This explanation is further supported by the observation that Multiscale Flow never overfits when trained with generative loss, and there is only slight overfitting when trained using hybrid loss with a large $\lambda$, which can be easily controlled with early stopping. The CNN training, on the other hand, overfits more easily due to its high sample complexity and requires more regularization techniques. % \cite{sharma2023inprep}. 
In the future, we plan to investigate this topic more thoroughly and perform a detailed comparison of the two approaches with varying training sizes.

The low asymptotic error of discriminative models and low sample complexity of generative models can be understood as a bias-variance trade-off. To achieve the optimal balance of the trade-off, several works have proposed building hybrid models \cite{raina2003classification,mccallum2006multi,bouchard2007bias,Liu2020hybrid}. Multiscale Flow is essentially a hybrid model, trained with a weighted combination of the generative loss (Equation \ref{eq:Lg}) and discriminative loss (Equation \ref{eq:Ld}). The interpolation parameter $\lambda$ balances the tradeoff of two approaches.

Apart from low sample complexity, another advantage of learning the likelihood function is robustness. The likelihood value itself contains information about whether the data may be contaminated by unknown systematic effects. As shown in the lower right panel of Figure \ref{fig:distribution_shift}, by comparing the likelihood value of a given data to those of the training data, we can tell whether the data is an outlier.
It has also been suggested that generative models and hybrid models are more robust to adversarial attacks \cite{Li2019generative,Liu2020hybrid}, which could bias the parameter inference \cite{horowitz2022plausible}. In the future, we plan to study more on making robust constraints against systematic effects.

\subsection*{Comparison with diffusion models}

Diffusion models have been shown to generate realistic astrophysical fields \cite{smith2022realistic,mudur2022can,zhao2023can}, and to achieve state-of-the-art performance on image density estimation tasks \cite{kingma2021variational}. However, they seem to have difficulty producing reliable posterior constraints \cite{cuesta2023diffusion}. After all, the posterior is determined by the difference of log-likelihood across different conditional parameters, not the averaged log-likelihood. It has been suggested that different metrics (e.g., well-calibrated posterior v.s. realistic samples) are largely independent of each other in high dimensions, and good performance on one criterion does not imply good performance on other criteria \cite{Theis2016}. In our experiments, we find that optimizing the model only with log-likelihood is not enough to produce reliable posteriors, due to the high asymptotic error of generative models. We train the model with hybrid loss to reduce the asymptotic error, which requires sampling the posterior during training with HMC. Considering that diffusion models are computationally too expensive to run HMC on the fly, we choose normalizing flows in this work.

\subsection*{Comparison with TRENF}
Translation and Rotation Equivariant Normalizing Flow has been shown to produce reliable and tight posterior constraints on Gaussian random field and mildly nonlinear matter density fields \cite{Dai2022a}. However, when we apply TRENF to weak lensing datasets in this work, it couldn't produce well-calibrated posteriors due to the restricted architecture. This motivates us to develop Multiscale Flow with affine coupling transforms \cite{dinh2016density,kingma2018glow}, which is able to approximate any probability distributions under mild conditions \cite{koehler2021representational}. Moreover, the multiscale decomposition of the likelihood enables scale-dependent posterior analysis that helps to detect domain shifts between training simulations and observed data.

\section{Discussion}

In this paper, we develop a Multiscale Flow model for field-level cosmological inference. Multiscale Flow tries to model the likelihood function of the cosmological field without any dimension reduction. If the field is learned perfectly, the resulting likelihood analysis becomes optimal. On mock weak lensing convergence dataset we demonstrate that the constraining power of Multiscale Flow outperforms the power spectrum in terms of Figure of Merit by factors of 2.5 - 4, depending on the noise level, and outperforms CNN  by a factor 2 for the most realistic case with a noise level of $20\mathrm{arcmin}^{-2}$ and with baryon marginalization. 

Multiscale Flow enables field-level scale-dependent posterior analysis, which helps the identification of scale-dependent systematics that are not accurately modeled in training simulations. We demonstrate that it is able to identify distribution shifts on weak lensing maps with baryonic physics if the model is trained with dark-matter-only maps.

In this paper, our main focus is optimal and robust field-level likelihood analysis, but we also show that Multiscale Flow can be used for fast sample generation and super-resolution, replacing 
the need for expensive N-body simulations and ray tracing. 
We expect many other applications of Multiscale Flow, such as 21cm and other intensity maps, weak lensing maps, projected galaxy clustering, X-ray and thermal SZ maps, etc. Multiscale Flow can also be used to model 3D galaxy fields or 1D spectrum data like Lyman alpha forest. 

Multiscale Flow can be generalized to model maps with multiple channels $x_{2^n} = \{x_{2^n}^c\}_{c=1}^C$, where $c$ represents the index of channels, and $C$ is the total number of channels. Here the channels could represent different tomographic bins of cosmic shear analysis, or different tracers on the same area of the sky, such as galaxies and weak lensing. We can still use Equation \ref{eq:decomposition} to decompose the likelihood of input maps with multiple channels, and each term can be further decomposed with
\begin{eqnarray}
     \log p(x_{2^k}|y) =& \sum_{c=1}^C \log p(x_{2^k}^c|x_{2^k}^1,\cdots,x_{2^k}^{c-1},y) ,\nonumber \\
     \log p(x^d_{2^m}|x_{2^m},y) =%& \nonumber\\ 
     \sum_{c=1}^C &\log p(x_{2^m}^{d,c}|x_{2^m}^{d,1},\cdots,x_{2^m}^{d,c-1},x_{2^m},y) ,\nonumber
\end{eqnarray}
which allows us to check for consistency between different channels.

Multiscale Flow can also be generalized to model maps with survey masks. Following the strategy developed in \cite{Dai2022a}, one can first sample noise at the masked region, and then introduce position-dependent flow transformation to the model to learn the effect of survey mask. It can thus be applied
to realistic surveys such as 
Hyper Suprime-Cam \cite{aihara2018hyper}, Euclid \cite{laureijs2011euclid}, or Vera C. Rubin Observatory Legacy Survey of Space and Time \cite{ivezic2019lsst} for their robust and optimal analysis. 

%%%%%%%%%%%% Supplementary Methods %%%%%%%%%%%%
%\footnotesize
\section*{Materials and Methods}

\subsection*{Dark-matter-only weak lensing maps}
\label{subsec:DMO_WL}

The weak lensing convergence maps from Gupta et al.\cite{Gupta2018a} are generated from a suite of 75 N-body simulations with spatially flat $\Lambda$CDM cosmologies. Each simulation differs in cosmological parameters $\Omega_m$ and $\sigma_8$, while the other cosmological parameters are fixed at $\Omega_b=0.046$, $h=0.72$ and $n_s=0.96$. The two cosmological parameters $\Omega_m$ and $\sigma_8$ are sampled non-uniformly with density increases towards $\Omega_m = 0.26$ and $\sigma_8 = 0.8$. Each simulation evolves $512^3$ dark matter particles in a $240 h^{-1} \mathrm{Mpc}$ box with N-body code gadget-2 \cite{springel2005cosmological}. A series of snapshots are saved between redshifts $0 < z < 1$ such that adjacent snapshots are separated by $80 h^{-1}\mathrm{Mpc}$ in comoving distance. 

Weak lensing convergence maps with field of view $3.5 \times 3.5$ $\mathrm{deg}^2$ are then generated by ray-traced the snapshots of N-body simulations to redshift $z=1$ with multiple lens plane algorithm \cite{Schneider1992gravitational}. 512 pseudo-independent maps are created from each simulation by randomly rotating, flipping, and shifting the simulation snapshots. We refer the reader to Gupta et al.\cite{Gupta2018a} for a detailed description of how these data were generated. 

Following Ribli et al.\cite{Ribli2019a}, we downsample the maps from resolution $1024^2$ to resolution $512^2$ ($\sim 0.4$ arcmin), and add Gaussian galaxy shape noise to the maps with a standard deviation
\begin{equation}
    \label{eq:noise}
    \sigma = \frac{\sigma_{\epsilon}}{\sqrt{2n_{\mathrm{gal}}A_{\mathrm{pixel}}}}, 
\end{equation}
where $\sigma_{\epsilon} = 0.4$ is the mean intrinsic ellipticity of galaxies and $A_{\mathrm{pixel}}$ the area of the pixel. For this dataset we consider three different galaxy densities: $n_g=10\ \mathrm{arcmin}^{-2}$, $n_g=30\ \mathrm{arcmin}^{-2}$ and $n_g=100\ \mathrm{arcmin}^{-2}$. Ribli et al.\cite{Ribli2019a} smooth the maps with a 1 arcmin Gaussian kernel to increase the signal-to-noise (S/N) ratio and removes the information at very small scales where baryonic physics alters the matter distribution. In our analysis, however, we do not smooth the noisy maps. This is because our Normalizing Flow models the likelihood function by mapping the convergence map to a Gaussian random field of the same dimensionality, implicitly assuming that the input map is full-ranked. With Gaussian smoothing, the small-scale modes of smoothed maps become degenerate and the probability distribution is no longer full-ranked, leading to model failure in our analysis.

\subsection*{Weak lensing maps with baryon}

\iffalse
\begin{table*}
  \caption{Free parameters of the baryonic correction model, \textcolor{red}{Copied from CNN paper}}
  \label{tab:BCM}
  \vskip 0.15in
  \centering
  \begin{tabular}{>{\centering}c>{\centering}c>{\centering}c>{\centering\arraybackslash}c}
    \toprule
    Parameter & Physical mearning & Fiducial value & Prior bound\\ 
    \midrule
    $M_c (h^{-1}M_{\odot})$ & characteristic halo mass for retaining half of the total gas & $3.3\times 10^{13} $ & $[5.9\times 10^{11}, 4.4\times 10^{15}]$ \\
    $M_{1,0} (h^{-1}M_{\odot})$ & characteristic halo mass for a galaxy mass fraction of $0.023$ & $8.63\times 10^{11} $ & $[9.3\times 10^{10}, 1.1\times 10^{13}]$ \\
    $\eta$ & maximum distance of the ejected gas from the parent halo & $0.54$ & $[0.12, 2.7]$ \\
    $\beta$ & logarithmic slope of the gas fraction vs. the halo mass & $0.12$ & $[0.026, 3.8]$ \\
    \bottomrule
  \end{tabular}
  \vskip -0.1in
\end{table*}
\fi

To study the impact of baryonic effects in our analysis, we also consider weak lensing convergence maps from Lu et al.\cite{Lu2022simultaneously}. These maps are generated from the same set of N-body simulations and ray-tracing algorithms as the dark-matter-only maps described above, and have the same resolution and field of view. The main difference is that the simulation snapshots are post-processed to include the baryonic effects. We briefly describe this post-processing step below and refer the reader to Lu et al.\cite{Lu2021impact,Lu2022simultaneously} for more details. 

Lu et al.\cite{Lu2022simultaneously} find all dark matter halos with mass $> 10^{12} M_{\odot}$ in the simulation snapshots, and replace the halo particles with spherically symmetric analytical halo profiles to characterize the matter distribution inside halos. The analytical halo profile is given by Baryon Correction Model \cite[BCM,][]{Aric02020a}, which describes the halos with four components: the central galaxy (stars), bounded gas, ejected gas (due to AGN feedback), and relaxed dark matter. The masses and profiles of these four components are parametrized by four free parameters: $M_c$ (the characteristic halo mass for retaining half of the total gas), $M_{1,0}$ (the characteristic halo mass for a galaxy mass fraction of $0.023$), $\eta$ (the maximum distance of the ejected gas from the parent halo), and $\beta$ (the logarithmic slope of the gas fraction vs. the halo mass). 
%The physical meanings, fiducial values, and prior bounds of these four baryonic parameters can be found in Table \ref{tab:BCM}. 
This post-processing step removes the substructure and non-spherical shape of the halos, but it has been shown that these morphological differences between the simulated halos and spherical analytical profiles are not statistically significant when compared to the uncertainties of the power spectrum and peak counts in an HSC-like survey \cite{Lu2021impact}. 

Lu et al.\cite{Lu2022simultaneously} create 2048 maps with different baryon parameters for each cosmology. They train CNN with the first 1024 maps, and use the other 1024 maps to measure the mean and covariance matrix of the learned statistics. In our analysis, we only use the first 1024 maps to train our Multiscale Flow and do not use the rest of the 1024 maps. 

Similar to the preprocessed steps of the dark-matter-only maps (described in the previous section), we downsample the maps to resolution $512^2$, and add Gaussian shape noise (Equation \ref{eq:noise}). For this dataset we consider four galaxy densities:  $n_g=10\ \mathrm{arcmin}^{-2}$, $n_g=20\ \mathrm{arcmin}^{-2}$, $n_g=50\ \mathrm{arcmin}^{-2}$ and $n_g=100\ \mathrm{arcmin}^{-2}$, to better compare our results with Lu et al.\cite{Lu2022simultaneously}.

\subsection*{Multiscale Flow Hyperparameters}
\label{subsec:MF_analysis}

%We decompose the weak lensing convergence maps into four different scales and apply Multiscale Flow to model their likelihood functions separately
%\begin{eqnarray}
%\label{eq:lnL_WL}
%    \log p(x_{512}|y) = &\log p(x_{64}|y) + \log p(x^d_{64}|x_{64},y) + \nonumber\\
%    &\log p(x^d_{128}|x_{128},y) + \log p(x^d_{256}|x_{256},y).
%\end{eqnarray}
We use $p=12$ block flows to model the large-scale term $\log p(x_{64}|y)$, and $q=4$ block flows to model each of the three small-scale terms. The CNN in Equation \ref{eq:affine1} is chosen to be a convolutional residual neural network with 2 residual blocks and 64 hidden channels in the residual blocks. 
%As described in Section \ref{subsec:training}, we firstly optimize the model by maximizing the data likelihood (Equation \ref{eq:Lg}), and then further tune the model by optimizing the posteriors with a KL divergence term to encourage conservative posteriors (Equation \ref{eq:Ldnew}).  

\subsection*{Summary Statistics Analysis}

In this paper, we compare the performance of Multiscale Flow with analysis based on summary statistics. We consider not only standard summary statistics such as power spectrum and peak count, but also novel statistics such as scattering transform and convolutional neural networks (CNNs).

\subsubsection*{Power Spectrum}
We compute the power spectrum of the convergence maps using the publicly available LensTools package \cite{Petri2016mocking}. The power spectrum is calculated in 20 bins in the range $100 \leq l \leq 37500$ with logarithmic spacing, following the settings adopted in Ribli et al.\cite{Ribli2019a} and Cheng et al.\cite{Cheng2020a}. We take the logarithm of the power spectrum to be observable for parameter inference.\\

\subsubsection*{Peak Count} 
{Peak count has been widely used in current weak lensing analysis \cite{martinet2018kids,harnois2021cosmic,zurcher2022dark,liu2023cosmological}. In Table \ref{tab:FoM}, we take the peak count measurement from Ribli et al. \cite{Ribli2019a}, who identify the local maxima of convergence maps and measure the binned histogram of the peaks as a function of their $\kappa$ value. They use 20 linearly spaced $\kappa$ bins in total.\\

\subsubsection*{Scattering transform}
Originally proposed by Mallat\cite{mallat2012group} as a tool to extract information from high-dimensional data, scattering transform has recently been applied to cosmological data analysis and shown improvement over the power spectrum in low noise regime \cite[e.g.,][]{Cheng2020a, allys2020new, cheng2021weak, valogiannis2022towards}. For a given input field, the scattering transform first generates a group of new fields by recursively applying wavelet convolutions and modulus. The expected values of these fields are then defined as the scattering coefficients and used as the summary statistics. In this paper we compare our results directly to Cheng et al.\cite{Cheng2020a}, who estimate the constraining power of scattering transform using Fisher forecast on the same dataset. \\

\subsubsection*{Convolutional Neural Networks (CNN)}
Several studies have explored using CNNs to construct summary statistics for cosmological inference \cite{fluri2018cosmological,Makinen2021a,Gupta2018a,jeffrey2021likelihood,Ribli2019a,Lu2022simultaneously,lu2023cosmological}. In this work we compare our results on dark-matter-only weak lensing maps with Ribli et al.\cite{Ribli2019a}, and compare our results with Lu et al.\cite{Lu2022simultaneously} on weak lensing maps with baryons. Ribli et al.\cite{Ribli2019a} and Lu et al.\cite{Lu2022simultaneously} train CNNs to predict cosmological parameters from the same convergence maps used in this work. Then they view these predicted parameters as summary statistics, and build Gaussian likelihood on these statistics for inference.

%%%%%%%%%%%%% Acknowledgements %%%%%%%%%%%%%
%\footnotesize
\section*{Acknowledgements}
We thank the Columbia Lensing group (\url{http://columbialensing.org}) for making their simulations available. B.D. thanks Xiangchong Li for helpful discussions on wavelet transform. This work is supported by U.S. Department of Energy, Office of Science, Office of Advanced Scientific Computing Research under Contract No. DE-AC02-05CH11231 at Lawrence Berkeley National Laboratory to enable research for Data-intensive Machine Learning and Analysis.

%%%%%%%%%%%%%%   Bibliography   %%%%%%%%%%%%%%
\normalsize
\bibliography{references}

\begin{thebibliography}{88}
\providecommand{\natexlab}[1]{#1}
\providecommand{\url}[1]{\texttt{#1}}
\expandafter\ifx\csname urlstyle\endcsname\relax
  \providecommand{\doi}[1]{doi: #1}\else
  \providecommand{\doi}{doi: \begingroup \urlstyle{rm}\Url}\fi

\bibitem[{Peebles}(1980)]{peebles1980large}
P.~J.~E. {Peebles}.
\newblock \emph{{The large-scale structure of the universe}}.
\newblock 1980.

\bibitem[{Peebles} and {Groth}(1975)]{peebles1975statistical}
P.~J.~E. {Peebles} and E.~J. {Groth}.
\newblock {Statistical analysis of catalogs of extragalactic objects. V.
  Three-point correlation function for the galaxy distribution in the Zwicky
  catalog.}
\newblock \emph{The Astrophysical Journal}, 196:\penalty0 1--11, February 1975.
\newblock \doi{10.1086/153390}.

\bibitem[Sefusatti et~al.(2006)Sefusatti, Crocce, Pueblas, and
  Scoccimarro]{sefusatti2006cosmology}
Emiliano Sefusatti, Mart{\'\i}n Crocce, Sebasti{\'a}n Pueblas, and Rom{\'a}n
  Scoccimarro.
\newblock Cosmology and the bispectrum.
\newblock \emph{Physical Review D}, 74\penalty0 (2):\penalty0 023522, 2006.

\bibitem[Semboloni et~al.(2011)Semboloni, Schrabback, van Waerbeke, Vafaei,
  Hartlap, and Hilbert]{semboloni2011weak}
Elisabetta Semboloni, Tim Schrabback, Ludovic van Waerbeke, Sanaz Vafaei, Jan
  Hartlap, and Stefan Hilbert.
\newblock Weak lensing from space: first cosmological constraints from
  three-point shear statistics.
\newblock \emph{Monthly Notices of the Royal Astronomical Society},
  410\penalty0 (1):\penalty0 143--160, 2011.

\bibitem[Fu et~al.(2014)Fu, Kilbinger, Erben, Heymans, Hildebrandt, Hoekstra,
  Kitching, Mellier, Miller, Semboloni, et~al.]{fu2014cfhtlens}
Liping Fu, Martin Kilbinger, Thomas Erben, Catherine Heymans, Hendrik
  Hildebrandt, Henk Hoekstra, Thomas~D Kitching, Yannick Mellier, Lance Miller,
  Elisabetta Semboloni, et~al.
\newblock Cfhtlens: cosmological constraints from a combination of cosmic shear
  two-point and three-point correlations.
\newblock \emph{Monthly Notices of the Royal Astronomical Society},
  441\penalty0 (3):\penalty0 2725--2743, 2014.

\bibitem[Kendall et~al.(1946)]{kendall1946advanced}
Maurice~George Kendall et~al.
\newblock The advanced theory of statistics.
\newblock \emph{The advanced theory of statistics.}, \penalty0 (2nd Ed), 1946.

\bibitem[{Neyrinck} et~al.(2009){Neyrinck}, {Szapudi}, and
  {Szalay}]{neyrinck2009rejuvenating}
Mark~C. {Neyrinck}, Istv{\'a}n {Szapudi}, and Alexander~S. {Szalay}.
\newblock {Rejuvenating the Matter Power Spectrum: Restoring Information with a
  Logarithmic Density Mapping}.
\newblock \emph{The Astrophysical Journal Letters}, 698\penalty0 (2):\penalty0
  L90--L93, June 2009.
\newblock \doi{10.1088/0004-637X/698/2/L90}.

\bibitem[White(2016)]{white2016marked}
Martin White.
\newblock A marked correlation function for constraining modified gravity
  models.
\newblock \emph{Journal of Cosmology and Astroparticle Physics}, 2016\penalty0
  (11):\penalty0 057, 2016.

\bibitem[{Jain} and {Van Waerbeke}(2000)]{Jain2000statistics}
Bhuvnesh {Jain} and Ludovic {Van Waerbeke}.
\newblock {Statistics of Dark Matter Halos from Gravitational Lensing}.
\newblock \emph{The Astrophysical Journal Letters}, 530\penalty0 (1):\penalty0
  L1--L4, February 2000.
\newblock \doi{10.1086/312480}.

\bibitem[{Kratochvil} et~al.(2010){Kratochvil}, {Haiman}, and
  {May}]{kratochvil2010probing}
Jan~M. {Kratochvil}, Zolt{\'a}n {Haiman}, and Morgan {May}.
\newblock {Probing cosmology with weak lensing peak counts}.
\newblock \emph{Physical Review D}, 81\penalty0 (4):\penalty0 043519, February
  2010.
\newblock \doi{10.1103/PhysRevD.81.043519}.

\bibitem[White(1979)]{white1979hierarchy}
Simon~DM White.
\newblock The hierarchy of correlation functions and its relation to other
  measures of galaxy clustering.
\newblock \emph{Monthly Notices of the Royal Astronomical Society},
  186\penalty0 (2):\penalty0 145--154, 1979.

\bibitem[{Pisani} et~al.(2019){Pisani}, {Massara}, {Spergel}, {Alonso},
  {Baker}, {Cai}, {Cautun}, {Davies}, {Demchenko}, {Dor{\'e}}, {Goulding},
  {Habouzit}, {Hamaus}, {Hawken}, {Hirata}, {Ho}, {Jain}, {Kreisch}, {Marulli},
  {Padilla}, {Pollina}, {Sahl{\'e}n}, {Sheth}, {Somerville}, {Szapudi}, {van de
  Weygaert}, {Villaescusa-Navarro}, {Wandelt}, and {Wang}]{pisani2019cosmic}
Alice {Pisani}, Elena {Massara}, David~N. {Spergel}, David {Alonso}, Tessa
  {Baker}, Yan-Chuan {Cai}, Marius {Cautun}, Christopher {Davies}, Vasiliy
  {Demchenko}, Olivier {Dor{\'e}}, Andy {Goulding}, M{\'e}lanie {Habouzit},
  Nico {Hamaus}, Adam {Hawken}, Christopher~M. {Hirata}, Shirley {Ho}, Bhuvnesh
  {Jain}, Christina~D. {Kreisch}, Federico {Marulli}, Nelson {Padilla}, Giorgia
  {Pollina}, Martin {Sahl{\'e}n}, Ravi~K. {Sheth}, Rachel {Somerville}, Istvan
  {Szapudi}, Rien {van de Weygaert}, Francisco {Villaescusa-Navarro},
  Benjamin~D. {Wandelt}, and Yun {Wang}.
\newblock {Cosmic voids: a novel probe to shed light on our Universe}.
\newblock \emph{Bulletin of the American Astronomical Society}, 51\penalty0
  (3):\penalty0 40, May 2019.
\newblock \doi{10.48550/arXiv.1903.05161}.

\bibitem[{Mecke} et~al.(1994){Mecke}, {Buchert}, and {Wagner}]{mecke1994robust}
K.~R. {Mecke}, T.~{Buchert}, and H.~{Wagner}.
\newblock {Robust morphological measures for large-scale structure in the
  Universe}.
\newblock \emph{Astronomy and Astrophysics}, 288:\penalty0 697--704, August
  1994.
\newblock \doi{10.48550/arXiv.astro-ph/9312028}.

\bibitem[{Cheng} et~al.(2020){Cheng}, {Ting}, {M{\'e}nard}, and
  {Bruna}]{Cheng2020a}
Sihao {Cheng}, Yuan-Sen {Ting}, Brice {M{\'e}nard}, and Joan {Bruna}.
\newblock {A new approach to observational cosmology using the scattering
  transform}.
\newblock \emph{Monthly Notices of the Royal Astronomical Society},
  499\penalty0 (4):\penalty0 5902--5914, December 2020.
\newblock \doi{10.1093/mnras/staa3165}.

\bibitem[Allys et~al.(2020)Allys, Marchand, Cardoso, Villaescusa-Navarro, Ho,
  and Mallat]{allys2020new}
Erwan Allys, T~Marchand, J-F Cardoso, F~Villaescusa-Navarro, S~Ho, and
  S~Mallat.
\newblock New interpretable statistics for large-scale structure analysis and
  generation.
\newblock \emph{Physical Review D}, 102\penalty0 (10):\penalty0 103506, 2020.

\bibitem[Fluri et~al.(2018)Fluri, Kacprzak, Refregier, Amara, Lucchi, and
  Hofmann]{fluri2018cosmological}
Janis Fluri, Tomasz Kacprzak, Alexandre Refregier, Adam Amara, Aurelien Lucchi,
  and Thomas Hofmann.
\newblock Cosmological constraints from noisy convergence maps through deep
  learning.
\newblock \emph{Physical Review D}, 98\penalty0 (12):\penalty0 123518, 2018.

\bibitem[{Charnock} et~al.(2018){Charnock}, {Lavaux}, and
  {Wandelt}]{Charnock18}
Tom {Charnock}, Guilhem {Lavaux}, and Benjamin~D. {Wandelt}.
\newblock {Automatic physical inference with information maximizing neural
  networks}.
\newblock \emph{Physical Review D}, 97\penalty0 (8):\penalty0 083004, April
  2018.
\newblock \doi{10.1103/PhysRevD.97.083004}.

\bibitem[{Makinen} et~al.(2021){Makinen}, {Charnock}, {Alsing}, and
  {Wandelt}]{Makinen2021a}
T.~Lucas {Makinen}, Tom {Charnock}, Justin {Alsing}, and Benjamin~D. {Wandelt}.
\newblock {Lossless, scalable implicit likelihood inference for cosmological
  fields}.
\newblock \emph{Journal of Cosmology and Astroparticle Physics}, 2021\penalty0
  (11):\penalty0 049, November 2021.
\newblock \doi{10.1088/1475-7516/2021/11/049}.

\bibitem[{Gupta} et~al.(2018){Gupta}, {Zorrilla Matilla}, {Hsu}, and
  {Haiman}]{Gupta2018a}
Arushi {Gupta}, Jos{\'e}~Manuel {Zorrilla Matilla}, Daniel {Hsu}, and
  Zolt{\'a}n {Haiman}.
\newblock {Non-Gaussian information from weak lensing data via deep learning}.
\newblock \emph{Physical Review D}, 97\penalty0 (10):\penalty0 103515, May
  2018.
\newblock \doi{10.1103/PhysRevD.97.103515}.

\bibitem[{Jeffrey} et~al.(2021){Jeffrey}, {Alsing}, and
  {Lanusse}]{jeffrey2021likelihood}
Niall {Jeffrey}, Justin {Alsing}, and Fran{\c{c}}ois {Lanusse}.
\newblock {Likelihood-free inference with neural compression of DES SV weak
  lensing map statistics}.
\newblock \emph{Monthly Notices of the Royal Astronomical Society},
  501\penalty0 (1):\penalty0 954--969, February 2021.
\newblock \doi{10.1093/mnras/staa3594}.

\bibitem[Cranmer et~al.(2020)Cranmer, Brehmer, and Louppe]{Cranmer2020a}
Kyle Cranmer, Johann Brehmer, and Gilles Louppe.
\newblock The frontier of simulation-based inference.
\newblock \emph{Proceedings of the National Academy of Sciences}, 117\penalty0
  (48):\penalty0 30055--30062, 2020.

\bibitem[{Jasche} and {Wandelt}(2013)]{Jasche2013a}
Jens {Jasche} and Benjamin~D. {Wandelt}.
\newblock {Bayesian physical reconstruction of initial conditions from
  large-scale structure surveys}.
\newblock \emph{Monthly Notices of the Royal Astronomical Society},
  432\penalty0 (2):\penalty0 894--913, June 2013.
\newblock \doi{10.1093/mnras/stt449}.

\bibitem[{Kitaura}(2013)]{Kitaura2013a}
F.~S. {Kitaura}.
\newblock {The initial conditions of the universe from constrained
  simulations.}
\newblock \emph{Monthly Notices of the Royal Astronomical Society},
  429:\penalty0 L84--L88, February 2013.
\newblock \doi{10.1093/mnrasl/sls029}.

\bibitem[{Wang} et~al.(2014){Wang}, {Mo}, {Yang}, {Jing}, and {Lin}]{Wang2014a}
Huiyuan {Wang}, H.~J. {Mo}, Xiaohu {Yang}, Y.~P. {Jing}, and W.~P. {Lin}.
\newblock {ELUCID{\textemdash}Exploring the Local Universe with the
  Reconstructed Initial Density Field. I. Hamiltonian Markov Chain Monte Carlo
  Method with Particle Mesh Dynamics}.
\newblock \emph{The Astrophysical Journal}, 794\penalty0 (1):\penalty0 94,
  October 2014.
\newblock \doi{10.1088/0004-637X/794/1/94}.

\bibitem[{Seljak} et~al.(2017){Seljak}, {Aslanyan}, {Feng}, and
  {Modi}]{Seljak2017a}
Uro{\v{s}} {Seljak}, Grigor {Aslanyan}, Yu~{Feng}, and Chirag {Modi}.
\newblock {Towards optimal extraction of cosmological information from
  nonlinear data}.
\newblock \emph{Journal of Cosmology and Astroparticle Physics}, 2017\penalty0
  (12):\penalty0 009, December 2017.
\newblock \doi{10.1088/1475-7516/2017/12/009}.

\bibitem[{Porqueres} et~al.(2022){Porqueres}, {Heavens}, {Mortlock}, and
  {Lavaux}]{2022Porqueres}
Natalia {Porqueres}, Alan {Heavens}, Daniel {Mortlock}, and Guilhem {Lavaux}.
\newblock {Lifting weak lensing degeneracies with a field-based likelihood}.
\newblock \emph{MNRAS}, 509\penalty0 (3):\penalty0 3194--3202, January 2022.
\newblock \doi{10.1093/mnras/stab3234}.

\bibitem[{Dai} and {Seljak}(2022)]{Dai2022a}
Biwei {Dai} and Uro{\v{s}} {Seljak}.
\newblock {Translation and rotation equivariant normalizing flow (TRENF) for
  optimal cosmological analysis}.
\newblock \emph{Monthly Notices of the Royal Astronomical Society},
  516\penalty0 (2):\penalty0 2363--2373, October 2022.
\newblock \doi{10.1093/mnras/stac2010}.

\bibitem[{Hassan} et~al.(2021){Hassan}, {Villaescusa-Navarro}, {Wandelt},
  {Spergel}, {Angl{\'e}s-Alc{\'a}zar}, {Genel}, {Cranmer}, {Bryan}, {Dav{\'e}},
  {Somerville}, {Eickenberg}, {Narayanan}, {Ho}, and
  {Andrianomena}]{Hassan2021}
Sultan {Hassan}, Francisco {Villaescusa-Navarro}, Benjamin {Wandelt}, David~N.
  {Spergel}, Daniel {Angl{\'e}s-Alc{\'a}zar}, Shy {Genel}, Miles {Cranmer},
  Greg~L. {Bryan}, Romeel {Dav{\'e}}, Rachel~S. {Somerville}, Michael
  {Eickenberg}, Desika {Narayanan}, Shirley {Ho}, and Sambatra {Andrianomena}.
\newblock {HIFlow: Generating Diverse HI Maps Conditioned on Cosmology using
  Normalizing Flow}.
\newblock \emph{arXiv e-prints}, art. arXiv:2110.02983, October 2021.

\bibitem[Friedman and Hassan(2022)]{friedman2022higlow}
Roy Friedman and Sultan Hassan.
\newblock Higlow: Conditional normalizing flows for high-fidelity hi map
  modeling.
\newblock \emph{arXiv preprint arXiv:2211.12724}, 2022.

\bibitem[Elahi et~al.(2016)Elahi, Knebe, Pearce, Power, Yepes, Cui, Cunnama,
  Kay, Sembolini, Beck, et~al.]{elahi2016nifty}
Pascal~J Elahi, Alexander Knebe, Frazer~R Pearce, Chris Power, Gustavo Yepes,
  Weiguang Cui, Daniel Cunnama, Scott~T Kay, Federico Sembolini, Alexander~M
  Beck, et~al.
\newblock nifty galaxy cluster simulations--iii. the similarity and diversity
  of galaxies and subhaloes.
\newblock \emph{Monthly Notices of the Royal Astronomical Society},
  458\penalty0 (1):\penalty0 1096--1116, 2016.

\bibitem[Huang et~al.(2019)Huang, Eifler, Mandelbaum, and
  Dodelson]{huang2019modelling}
Hung-Jin Huang, Tim Eifler, Rachel Mandelbaum, and Scott Dodelson.
\newblock Modelling baryonic physics in future weak lensing surveys.
\newblock \emph{Monthly Notices of the Royal Astronomical Society},
  488\penalty0 (2):\penalty0 1652--1678, 2019.

\bibitem[Villaescusa-Navarro et~al.(2021)Villaescusa-Navarro,
  Angl{\'e}s-Alc{\'a}zar, Genel, Spergel, Li, Wandelt, Nicola, Thiele, Hassan,
  Matilla, et~al.]{villaescusa2021multifield}
Francisco Villaescusa-Navarro, Daniel Angl{\'e}s-Alc{\'a}zar, Shy Genel,
  David~N Spergel, Yin Li, Benjamin Wandelt, Andrina Nicola, Leander Thiele,
  Sultan Hassan, Jose Manuel~Zorrilla Matilla, et~al.
\newblock Multifield cosmology with artificial intelligence.
\newblock \emph{arXiv preprint arXiv:2109.09747}, 2021.

\bibitem[Pillepich et~al.(2018)Pillepich, Springel, Nelson, Genel, Naiman,
  Pakmor, Hernquist, Torrey, Vogelsberger, Weinberger,
  et~al.]{pillepich2018simulating}
Annalisa Pillepich, Volker Springel, Dylan Nelson, Shy Genel, Jill Naiman,
  R{\"u}diger Pakmor, Lars Hernquist, Paul Torrey, Mark Vogelsberger, Rainer
  Weinberger, et~al.
\newblock Simulating galaxy formation with the illustristng model.
\newblock \emph{Monthly Notices of the Royal Astronomical Society},
  473\penalty0 (3):\penalty0 4077--4106, 2018.

\bibitem[Dav{\'e} et~al.(2019)Dav{\'e}, Angl{\'e}s-Alc{\'a}zar, Narayanan, Li,
  Rafieferantsoa, and Appleby]{dave2019simba}
Romeel Dav{\'e}, Daniel Angl{\'e}s-Alc{\'a}zar, Desika Narayanan, Qi~Li, Mika~H
  Rafieferantsoa, and Sarah Appleby.
\newblock Simba: Cosmological simulations with black hole growth and feedback.
\newblock \emph{Monthly Notices of the Royal Astronomical Society},
  486\penalty0 (2):\penalty0 2827--2849, 2019.

\bibitem[{Krause} et~al.(2017){Krause}, {Eifler}, {Zuntz}, {Friedrich},
  {Troxel}, {Dodelson}, {Blazek}, {Secco}, {MacCrann}, {Baxter}, {Chang},
  {Chen}, {Crocce}, {DeRose}, {Ferte}, {Kokron}, {Lacasa}, {Miranda}, {Omori},
  {Porredon}, {Rosenfeld}, {Samuroff}, {Wang}, {Wechsler}, {Abbott}, {Abdalla},
  {Allam}, {Annis}, {Bechtol}, {Benoit-Levy}, {Bernstein}, {Brooks}, {Burke},
  {Capozzi}, {Carrasco Kind}, {Carretero}, {D'Andrea}, {da Costa}, {Davis},
  {DePoy}, {Desai}, {Diehl}, {Dietrich}, {Evrard}, {Flaugher}, {Fosalba},
  {Frieman}, {Garcia-Bellido}, {Gaztanaga}, {Giannantonio}, {Gruen}, {Gruendl},
  {Gschwend}, {Gutierrez}, {Honscheid}, {James}, {Jeltema}, {Kuehn},
  {Kuhlmann}, {Lahav}, {Lima}, {Maia}, {March}, {Marshall}, {Martini},
  {Menanteau}, {Miquel}, {Nichol}, {Plazas}, {Romer}, {Rykoff}, {Sanchez},
  {Scarpine}, {Schindler}, {Schubnell}, {Sevilla-Noarbe}, {Smith},
  {Soares-Santos}, {Sobreira}, {Suchyta}, {Swanson}, {Tarle}, {Tucker},
  {Vikram}, {Walker}, and {Weller}]{Krause2017dark}
E.~{Krause}, T.~F. {Eifler}, J.~{Zuntz}, O.~{Friedrich}, M.~A. {Troxel},
  S.~{Dodelson}, J.~{Blazek}, L.~F. {Secco}, N.~{MacCrann}, E.~{Baxter},
  C.~{Chang}, N.~{Chen}, M.~{Crocce}, J.~{DeRose}, A.~{Ferte}, N.~{Kokron},
  F.~{Lacasa}, V.~{Miranda}, Y.~{Omori}, A.~{Porredon}, R.~{Rosenfeld},
  S.~{Samuroff}, M.~{Wang}, R.~H. {Wechsler}, T.~M.~C. {Abbott}, F.~B.
  {Abdalla}, S.~{Allam}, J.~{Annis}, K.~{Bechtol}, A.~{Benoit-Levy}, G.~M.
  {Bernstein}, D.~{Brooks}, D.~L. {Burke}, D.~{Capozzi}, M.~{Carrasco Kind},
  J.~{Carretero}, C.~B. {D'Andrea}, L.~N. {da Costa}, C.~{Davis}, D.~L.
  {DePoy}, S.~{Desai}, H.~T. {Diehl}, J.~P. {Dietrich}, A.~E. {Evrard},
  B.~{Flaugher}, P.~{Fosalba}, J.~{Frieman}, J.~{Garcia-Bellido},
  E.~{Gaztanaga}, T.~{Giannantonio}, D.~{Gruen}, R.~A. {Gruendl},
  J.~{Gschwend}, G.~{Gutierrez}, K.~{Honscheid}, D.~J. {James}, T.~{Jeltema},
  K.~{Kuehn}, S.~{Kuhlmann}, O.~{Lahav}, M.~{Lima}, M.~A.~G. {Maia},
  M.~{March}, J.~L. {Marshall}, P.~{Martini}, F.~{Menanteau}, R.~{Miquel},
  R.~C. {Nichol}, A.~A. {Plazas}, A.~K. {Romer}, E.~S. {Rykoff}, E.~{Sanchez},
  V.~{Scarpine}, R.~{Schindler}, M.~{Schubnell}, I.~{Sevilla-Noarbe},
  M.~{Smith}, M.~{Soares-Santos}, F.~{Sobreira}, E.~{Suchyta}, M.~E.~C.
  {Swanson}, G.~{Tarle}, D.~L. {Tucker}, V.~{Vikram}, A.~R. {Walker}, and
  J.~{Weller}.
\newblock {Dark Energy Survey Year 1 Results: Multi-Probe Methodology and
  Simulated Likelihood Analyses}.
\newblock \emph{arXiv e-prints}, art. arXiv:1706.09359, June 2017.

\bibitem[Taylor et~al.(2018)Taylor, Bernardeau, and Kitching]{taylor2018k}
Peter~L Taylor, Francis Bernardeau, and Thomas~D Kitching.
\newblock k-cut cosmic shear: Tunable power spectrum sensitivity to test
  gravity.
\newblock \emph{Physical Review D}, 98\penalty0 (8):\penalty0 083514, 2018.

\bibitem[{Krause} et~al.(2021){Krause}, {Fang}, {Pandey}, {Secco}, {Alves},
  {Huang}, {Blazek}, {Prat}, {Zuntz}, {Eifler}, {MacCrann}, {DeRose}, {Crocce},
  {Porredon}, {Jain}, {Troxel}, {Dodelson}, {Huterer}, {Liddle}, {Leonard},
  {Amon}, {Chen}, {Elvin-Poole}, {Fert{\'e}}, {Muir}, {Park}, {Samuroff},
  {Brandao-Souza}, {Weaverdyck}, {Zacharegkas}, {Rosenfeld}, {Campos},
  {Chintalapati}, {Choi}, {Di Valentino}, {Doux}, {Herner}, {Lemos},
  {Mena-Fern{\'a}ndez}, {Omori}, {Paterno}, {Rodriguez-Monroy}, {Rogozenski},
  {Rollins}, {Troja}, {Tutusaus}, {Wechsler}, {Abbott}, {Aguena}, {Allam},
  {Andrade-Oliveira}, {Annis}, {Bacon}, {Baxter}, {Bechtol}, {Bernstein},
  {Brooks}, {Buckley-Geer}, {Burke}, {Carnero Rosell}, {Carrasco Kind},
  {Carretero}, {Castander}, {Cawthon}, {Chang}, {Costanzi}, {da Costa},
  {Pereira}, {De Vicente}, {Desai}, {Diehl}, {Doel}, {Everett}, {Evrard},
  {Ferrero}, {Flaugher}, {Fosalba}, {Frieman}, {Garc{\'\i}a-Bellido},
  {Gaztanaga}, {Gerdes}, {Giannantonio}, {Gruen}, {Gruendl}, {Gschwend},
  {Gutierrez}, {Hartley}, {Hinton}, {Hollowood}, {Honscheid}, {Hoyle}, {Huff},
  {James}, {Kuehn}, {Kuropatkin}, {Lahav}, {Lima}, {Maia}, {Marshall},
  {Martini}, {Melchior}, {Menanteau}, {Miquel}, {Mohr}, {Morgan}, {Myles},
  {Palmese}, {Paz-Chinch{\'o}n}, {Petravick}, {Pieres}, {Plazas Malag{\'o}n},
  {Sanchez}, {Scarpine}, {Schubnell}, {Serrano}, {Sevilla-Noarbe}, {Smith},
  {Soares-Santos}, {Suchyta}, {Tarle}, {Thomas}, {To}, {Varga}, and
  {Weller}]{Krause2021dark}
E.~{Krause}, X.~{Fang}, S.~{Pandey}, L.~F. {Secco}, O.~{Alves}, H.~{Huang},
  J.~{Blazek}, J.~{Prat}, J.~{Zuntz}, T.~F. {Eifler}, N.~{MacCrann},
  J.~{DeRose}, M.~{Crocce}, A.~{Porredon}, B.~{Jain}, M.~A. {Troxel},
  S.~{Dodelson}, D.~{Huterer}, A.~R. {Liddle}, C.~D. {Leonard}, A.~{Amon},
  A.~{Chen}, J.~{Elvin-Poole}, A.~{Fert{\'e}}, J.~{Muir}, Y.~{Park},
  S.~{Samuroff}, A.~{Brandao-Souza}, N.~{Weaverdyck}, G.~{Zacharegkas},
  R.~{Rosenfeld}, A.~{Campos}, P.~{Chintalapati}, A.~{Choi}, E.~{Di Valentino},
  C.~{Doux}, K.~{Herner}, P.~{Lemos}, J.~{Mena-Fern{\'a}ndez}, Y.~{Omori},
  M.~{Paterno}, M.~{Rodriguez-Monroy}, P.~{Rogozenski}, R.~P. {Rollins},
  A.~{Troja}, I.~{Tutusaus}, R.~H. {Wechsler}, T.~M.~C. {Abbott}, M.~{Aguena},
  S.~{Allam}, F.~{Andrade-Oliveira}, J.~{Annis}, D.~{Bacon}, E.~{Baxter},
  K.~{Bechtol}, G.~M. {Bernstein}, D.~{Brooks}, E.~{Buckley-Geer}, D.~L.
  {Burke}, A.~{Carnero Rosell}, M.~{Carrasco Kind}, J.~{Carretero}, F.~J.
  {Castander}, R.~{Cawthon}, C.~{Chang}, M.~{Costanzi}, L.~N. {da Costa},
  M.~E.~S. {Pereira}, J.~{De Vicente}, S.~{Desai}, H.~T. {Diehl}, P.~{Doel},
  S.~{Everett}, A.~E. {Evrard}, I.~{Ferrero}, B.~{Flaugher}, P.~{Fosalba},
  J.~{Frieman}, J.~{Garc{\'\i}a-Bellido}, E.~{Gaztanaga}, D.~W. {Gerdes},
  T.~{Giannantonio}, D.~{Gruen}, R.~A. {Gruendl}, J.~{Gschwend},
  G.~{Gutierrez}, W.~G. {Hartley}, S.~R. {Hinton}, D.~L. {Hollowood},
  K.~{Honscheid}, B.~{Hoyle}, E.~M. {Huff}, D.~J. {James}, K.~{Kuehn},
  N.~{Kuropatkin}, O.~{Lahav}, M.~{Lima}, M.~A.~G. {Maia}, J.~L. {Marshall},
  P.~{Martini}, P.~{Melchior}, F.~{Menanteau}, R.~{Miquel}, J.~J. {Mohr},
  R.~{Morgan}, J.~{Myles}, A.~{Palmese}, F.~{Paz-Chinch{\'o}n}, D.~{Petravick},
  A.~{Pieres}, A.~A. {Plazas Malag{\'o}n}, E.~{Sanchez}, V.~{Scarpine},
  M.~{Schubnell}, S.~{Serrano}, I.~{Sevilla-Noarbe}, M.~{Smith},
  M.~{Soares-Santos}, E.~{Suchyta}, G.~{Tarle}, D.~{Thomas}, C.~{To}, T.~N.
  {Varga}, and J.~{Weller}.
\newblock {Dark Energy Survey Year 3 Results: Multi-Probe Modeling Strategy and
  Validation}.
\newblock \emph{arXiv e-prints}, art. arXiv:2105.13548, May 2021.

\bibitem[Doux et~al.(2021)Doux, Baxter, Lemos, Chang, Alarcon, Amon, Campos,
  Choi, Gatti, Gruen, et~al.]{doux2021dark}
Cyrille Doux, E~Baxter, Pablo Lemos, C~Chang, A~Alarcon, A~Amon, A~Campos,
  A~Choi, M~Gatti, D~Gruen, et~al.
\newblock Dark energy survey internal consistency tests of the joint
  cosmological probes analysis with posterior predictive distributions.
\newblock \emph{Monthly Notices of the Royal Astronomical Society},
  503\penalty0 (2):\penalty0 2688--2705, 2021.

\bibitem[Mallat(1989)]{mallat1989theory}
Stephane~G Mallat.
\newblock A theory for multiresolution signal decomposition: the wavelet
  representation.
\newblock \emph{IEEE transactions on pattern analysis and machine
  intelligence}, 11\penalty0 (7):\penalty0 674--693, 1989.

\bibitem[Starck and Murtagh(2007)]{starck2007astronomical}
J-L Starck and Fionn Murtagh.
\newblock \emph{Astronomical image and data analysis}.
\newblock Springer Science \& Business Media, 2007.

\bibitem[Haar(1910)]{Haar1910theorie}
Alfred Haar.
\newblock Zur theorie der orthogonalen funktionensysteme.
\newblock \emph{Mathematische Annalen.}, 69\penalty0 (3):\penalty0 331--371,
  1910.
\newblock ISSN 0025-5831.

\bibitem[Daubechies(1988)]{daubechies1988orthonormal}
Ingrid Daubechies.
\newblock Orthonormal bases of compactly supported wavelets.
\newblock \emph{Communications on pure and applied mathematics}, 41\penalty0
  (7):\penalty0 909--996, 1988.

\bibitem[Dinh et~al.(2017)Dinh, Sohl{-}Dickstein, and Bengio]{dinh2016density}
Laurent Dinh, Jascha Sohl{-}Dickstein, and Samy Bengio.
\newblock Density estimation using real {NVP}.
\newblock In \emph{5th International Conference on Learning Representations,
  {ICLR} 2017, Toulon, France, April 24-26, 2017, Conference Track
  Proceedings}. OpenReview.net, 2017.
\newblock URL \url{https://openreview.net/forum?id=HkpbnH9lx}.

\bibitem[Papamakarios et~al.(2017)Papamakarios, Murray, and
  Pavlakou]{papamakarios2017masked}
George Papamakarios, Iain Murray, and Theo Pavlakou.
\newblock Masked autoregressive flow for density estimation.
\newblock In Isabelle Guyon, Ulrike von Luxburg, Samy Bengio, Hanna~M. Wallach,
  Rob Fergus, S.~V.~N. Vishwanathan, and Roman Garnett, editors, \emph{Advances
  in Neural Information Processing Systems 30: Annual Conference on Neural
  Information Processing Systems 2017, December 4-9, 2017, Long Beach, CA,
  {USA}}, pages 2338--2347, 2017.

\bibitem[Kingma and Dhariwal(2018)]{kingma2018glow}
Diederik~P. Kingma and Prafulla Dhariwal.
\newblock Glow: Generative flow with invertible 1x1 convolutions.
\newblock In Samy Bengio, Hanna~M. Wallach, Hugo Larochelle, Kristen Grauman,
  Nicol{\`{o}} Cesa{-}Bianchi, and Roman Garnett, editors, \emph{Advances in
  Neural Information Processing Systems 31: Annual Conference on Neural
  Information Processing Systems 2018, NeurIPS 2018, December 3-8, 2018,
  Montr{\'{e}}al, Canada}, pages 10236--10245, 2018.

\bibitem[Yu et~al.(2020)Yu, Derpanis, and Brubaker]{yu2020wavelet}
Jason~J Yu, Konstantinos~G Derpanis, and Marcus~A Brubaker.
\newblock Wavelet flow: Fast training of high resolution normalizing flows.
\newblock \emph{Advances in Neural Information Processing Systems},
  33:\penalty0 6184--6196, 2020.

\bibitem[Ioffe and Szegedy(2015)]{ioffe2015batch}
Sergey Ioffe and Christian Szegedy.
\newblock Batch normalization: Accelerating deep network training by reducing
  internal covariate shift.
\newblock In \emph{International conference on machine learning}, pages
  448--456. PMLR, 2015.

\bibitem[Song and Kingma(2021)]{song2021train}
Yang Song and Diederik~P Kingma.
\newblock How to train your energy-based models.
\newblock \emph{arXiv preprint arXiv:2101.03288}, 2021.

\bibitem[Duane et~al.(1987)Duane, Kennedy, Pendleton, and
  Roweth]{duane1987hybrid}
Simon Duane, Anthony~D Kennedy, Brian~J Pendleton, and Duncan Roweth.
\newblock Hybrid monte carlo.
\newblock \emph{Physics letters B}, 195\penalty0 (2):\penalty0 216--222, 1987.

\bibitem[Tieleman(2008)]{tieleman2008training}
Tijmen Tieleman.
\newblock Training restricted boltzmann machines using approximations to the
  likelihood gradient.
\newblock In \emph{Proceedings of the 25th international conference on Machine
  learning}, pages 1064--1071, 2008.

\bibitem[Park et~al.(2022)Park, Allys, Villaescusa-Navarro, and
  Finkbeiner]{park2022quantification}
Core~Francisco Park, Erwan Allys, Francisco Villaescusa-Navarro, and Douglas~P
  Finkbeiner.
\newblock Quantification of high dimensional non-gaussianities and its
  implication to fisher analysis in cosmology.
\newblock \emph{arXiv preprint arXiv:2204.05435}, 2022.

\bibitem[{Ribli} et~al.(2019){Ribli}, {Pataki}, {Zorrilla Matilla}, {Hsu},
  {Haiman}, and {Csabai}]{Ribli2019a}
Dezs{\H{o}} {Ribli}, B{\'a}lint~{\'A}rmin {Pataki}, Jos{\'e}~Manuel {Zorrilla
  Matilla}, Daniel {Hsu}, Zolt{\'a}n {Haiman}, and Istv{\'a}n {Csabai}.
\newblock {Weak lensing cosmology with convolutional neural networks on noisy
  data}.
\newblock \emph{Monthly Notices of the Royal Astronomical Society},
  490\penalty0 (2):\penalty0 1843--1860, December 2019.
\newblock \doi{10.1093/mnras/stz2610}.

\bibitem[{Lu} et~al.(2022){Lu}, {Haiman}, and {Zorrilla
  Matilla}]{Lu2022simultaneously}
Tianhuan {Lu}, Zolt{\'a}n {Haiman}, and Jos{\'e}~Manuel {Zorrilla Matilla}.
\newblock {Simultaneously constraining cosmology and baryonic physics via deep
  learning from weak lensing}.
\newblock \emph{Monthly Notices of the Royal Astronomical Society},
  511\penalty0 (1):\penalty0 1518--1528, March 2022.
\newblock \doi{10.1093/mnras/stac161}.

\bibitem[{Aric{\`o}} et~al.(2020){Aric{\`o}}, {Angulo},
  {Hern{\'a}ndez-Monteagudo}, {Contreras}, {Zennaro}, {Pellejero-Iba{\~n}ez},
  and {Rosas-Guevara}]{Aric02020a}
Giovanni {Aric{\`o}}, Raul~E. {Angulo}, Carlos {Hern{\'a}ndez-Monteagudo},
  Sergio {Contreras}, Matteo {Zennaro}, Marcos {Pellejero-Iba{\~n}ez}, and
  Yetli {Rosas-Guevara}.
\newblock {Modelling the large-scale mass density field of the universe as a
  function of cosmology and baryonic physics}.
\newblock \emph{Monthly Notices of the Royal Astronomical Society},
  495\penalty0 (4):\penalty0 4800--4819, July 2020.
\newblock \doi{10.1093/mnras/staa1478}.

\bibitem[Villanueva-Domingo and
  Villaescusa-Navarro(2022)]{villanueva2022learning}
Pablo Villanueva-Domingo and Francisco Villaescusa-Navarro.
\newblock Learning cosmology and clustering with cosmic graphs.
\newblock \emph{The Astrophysical Journal}, 937\penalty0 (2):\penalty0 115,
  2022.

\bibitem[Fluri et~al.(2019)Fluri, Kacprzak, Lucchi, Refregier, Amara, Hofmann,
  and Schneider]{fluri2019cosmological}
Janis Fluri, Tomasz Kacprzak, Aurelien Lucchi, Alexandre Refregier, Adam Amara,
  Thomas Hofmann, and Aurel Schneider.
\newblock Cosmological constraints with deep learning from kids-450 weak
  lensing maps.
\newblock \emph{Physical Review D}, 100\penalty0 (6):\penalty0 063514, 2019.

\bibitem[Lu et~al.(2023)Lu, Haiman, and Li]{lu2023cosmological}
Tianhuan Lu, Zolt{\'a}n Haiman, and Xiangchong Li.
\newblock Cosmological constraints from hsc survey first-year data using deep
  learning.
\newblock \emph{arXiv preprint arXiv:2301.01354}, 2023.

\bibitem[Fluri et~al.(2022)Fluri, Kacprzak, Lucchi, Schneider, Refregier, and
  Hofmann]{fluri2022full}
Janis Fluri, Tomasz Kacprzak, Aurelien Lucchi, Aurel Schneider, Alexandre
  Refregier, and Thomas Hofmann.
\newblock Full w cdm analysis of kids-1000 weak lensing maps using deep
  learning.
\newblock \emph{Physical Review D}, 105\penalty0 (8):\penalty0 083518, 2022.

\bibitem[Ng and Jordan(2002)]{ng2002discriminative}
Andrew~Y Ng and Michael~I Jordan.
\newblock On discriminative vs. generative classifiers: A comparison of
  logistic regression and naive bayes.
\newblock In \emph{Advances in neural information processing systems}, pages
  841--848, 2002.

\bibitem[Yogatama et~al.(2017)Yogatama, Dyer, Ling, and
  Blunsom]{Yogatama2017generative}
Dani Yogatama, Chris Dyer, Wang Ling, and Phil Blunsom.
\newblock Generative and discriminative text classification with recurrent
  neural networks.
\newblock \emph{CoRR}, abs/1703.01898, 2017.
\newblock URL \url{http://arxiv.org/abs/1703.01898}.

\bibitem[Zheng et~al.(2023)Zheng, Wu, Bao, Cao, Li, and
  Zhu]{zheng2023revisiting}
Chenyu Zheng, Guoqiang Wu, Fan Bao, Yue Cao, Chongxuan Li, and Jun Zhu.
\newblock Revisiting discriminative vs. generative classifiers: Theory and
  implications.
\newblock In Andreas Krause, Emma Brunskill, Kyunghyun Cho, Barbara Engelhardt,
  Sivan Sabato, and Jonathan Scarlett, editors, \emph{International Conference
  on Machine Learning, {ICML} 2023, 23-29 July 2023, Honolulu, Hawaii, {USA}},
  volume 202 of \emph{Proceedings of Machine Learning Research}, pages
  42420--42477. {PMLR}, 2023.
\newblock URL \url{https://proceedings.mlr.press/v202/zheng23f.html}.

\bibitem[Raina et~al.(2003)Raina, Shen, Mccallum, and
  Ng]{raina2003classification}
Rajat Raina, Yirong Shen, Andrew Mccallum, and Andrew Ng.
\newblock Classification with hybrid generative/discriminative models.
\newblock \emph{Advances in neural information processing systems}, 16, 2003.

\bibitem[McCallum et~al.(2006)McCallum, Pal, Druck, and
  Wang]{mccallum2006multi}
Andrew McCallum, Chris Pal, Gregory Druck, and Xuerui Wang.
\newblock Multi-conditional learning: Generative/discriminative training for
  clustering and classification.
\newblock In \emph{AAAI}, volume~1, page~6, 2006.

\bibitem[Bouchard(2007)]{bouchard2007bias}
Guillaume Bouchard.
\newblock Bias-variance tradeoff in hybrid generative-discriminative models.
\newblock In \emph{Sixth International Conference on Machine Learning and
  Applications (ICMLA 2007)}, pages 124--129. IEEE, 2007.

\bibitem[Liu and Abbeel(2020)]{Liu2020hybrid}
Hao Liu and Pieter Abbeel.
\newblock Hybrid discriminative-generative training via contrastive learning.
\newblock \emph{CoRR}, abs/2007.09070, 2020.
\newblock URL \url{https://arxiv.org/abs/2007.09070}.

\bibitem[Li et~al.(2019)Li, Bradshaw, and Sharma]{Li2019generative}
Yingzhen Li, John Bradshaw, and Yash Sharma.
\newblock Are generative classifiers more robust to adversarial attacks?
\newblock In Kamalika Chaudhuri and Ruslan Salakhutdinov, editors,
  \emph{Proceedings of the 36th International Conference on Machine Learning,
  {ICML} 2019, 9-15 June 2019, Long Beach, California, {USA}}, volume~97 of
  \emph{Proceedings of Machine Learning Research}, pages 3804--3814. {PMLR},
  2019.
\newblock URL \url{http://proceedings.mlr.press/v97/li19a.html}.

\bibitem[Horowitz and Melchior(2022)]{horowitz2022plausible}
Benjamin Horowitz and Peter Melchior.
\newblock Plausible adversarial attacks on direct parameter inference models in
  astrophysics.
\newblock \emph{arXiv preprint arXiv:2211.14788}, 2022.

\bibitem[Smith et~al.(2022)Smith, Geach, Jackson, Arora, Stone, and
  Courteau]{smith2022realistic}
Michael~J Smith, James~E Geach, Ryan~A Jackson, Nikhil Arora, Connor Stone, and
  St{\'e}phane Courteau.
\newblock Realistic galaxy image simulation via score-based generative models.
\newblock \emph{Monthly Notices of the Royal Astronomical Society},
  511\penalty0 (2):\penalty0 1808--1818, 2022.

\bibitem[Mudur and Finkbeiner(2022)]{mudur2022can}
Nayantara Mudur and Douglas~P Finkbeiner.
\newblock Can denoising diffusion probabilistic models generate realistic
  astrophysical fields?
\newblock \emph{arXiv preprint arXiv:2211.12444}, 2022.

\bibitem[Zhao et~al.(2023)Zhao, Ting, Diao, and Mao]{zhao2023can}
Xiaosheng Zhao, Yuan-Sen Ting, Kangning Diao, and Yi~Mao.
\newblock Can diffusion model conditionally generate astrophysical images?
\newblock \emph{arXiv preprint arXiv:2307.09568}, 2023.

\bibitem[Kingma et~al.(2021)Kingma, Salimans, Poole, and
  Ho]{kingma2021variational}
Diederik Kingma, Tim Salimans, Ben Poole, and Jonathan Ho.
\newblock Variational diffusion models.
\newblock In M.~Ranzato, A.~Beygelzimer, Y.~Dauphin, P.S. Liang, and J.~Wortman
  Vaughan, editors, \emph{Advances in Neural Information Processing Systems},
  volume~34, pages 21696--21707. Curran Associates, Inc., 2021.
\newblock URL
  \url{https://proceedings.neurips.cc/paper_files/paper/2021/file/b578f2a52a0229873fefc2a4b06377fa-Paper.pdf}.

\bibitem[Cuesta-Lazaro and Mishra-Sharma(2023)]{cuesta2023diffusion}
Carolina Cuesta-Lazaro and Siddharth Mishra-Sharma.
\newblock Diffusion generative modeling for galaxy surveys: emulating
  clustering for inference at the field level.
\newblock 2023.

\bibitem[Theis et~al.(2016)Theis, van~den Oord, and Bethge]{Theis2016}
Lucas Theis, A{\"{a}}ron van~den Oord, and Matthias Bethge.
\newblock A note on the evaluation of generative models.
\newblock In Yoshua Bengio and Yann LeCun, editors, \emph{4th International
  Conference on Learning Representations, {ICLR} 2016, San Juan, Puerto Rico,
  May 2-4, 2016, Conference Track Proceedings}, 2016.
\newblock URL \url{http://arxiv.org/abs/1511.01844}.

\bibitem[Koehler et~al.(2021)Koehler, Mehta, and
  Risteski]{koehler2021representational}
Frederic Koehler, Viraj Mehta, and Andrej Risteski.
\newblock Representational aspects of depth and conditioning in normalizing
  flows.
\newblock In \emph{International Conference on Machine Learning}, pages
  5628--5636. PMLR, 2021.

\bibitem[Aihara et~al.(2018)Aihara, Arimoto, Armstrong, Arnouts, Bahcall,
  Bickerton, Bosch, Bundy, Capak, Chan, et~al.]{aihara2018hyper}
Hiroaki Aihara, Nobuo Arimoto, Robert Armstrong, St{\'e}phane Arnouts, Neta~A
  Bahcall, Steven Bickerton, James Bosch, Kevin Bundy, Peter~L Capak, James~HH
  Chan, et~al.
\newblock The hyper suprime-cam ssp survey: overview and survey design.
\newblock \emph{Publications of the Astronomical Society of Japan}, 70\penalty0
  (SP1):\penalty0 S4, 2018.

\bibitem[Laureijs et~al.(2011)Laureijs, Amiaux, Arduini, Augueres, Brinchmann,
  Cole, Cropper, Dabin, Duvet, Ealet, et~al.]{laureijs2011euclid}
Rene Laureijs, J~Amiaux, S~Arduini, J-L Augueres, J~Brinchmann, R~Cole,
  M~Cropper, C~Dabin, L~Duvet, A~Ealet, et~al.
\newblock Euclid definition study report.
\newblock \emph{arXiv preprint arXiv:1110.3193}, 2011.

\bibitem[Ivezi{\'c} et~al.(2019)Ivezi{\'c}, Kahn, Tyson, Abel, Acosta, Allsman,
  Alonso, AlSayyad, Anderson, Andrew, et~al.]{ivezic2019lsst}
{\v{Z}}eljko Ivezi{\'c}, Steven~M Kahn, J~Anthony Tyson, Bob Abel, Emily
  Acosta, Robyn Allsman, David Alonso, Yusra AlSayyad, Scott~F Anderson, John
  Andrew, et~al.
\newblock Lsst: from science drivers to reference design and anticipated data
  products.
\newblock \emph{The Astrophysical Journal}, 873\penalty0 (2):\penalty0 111,
  2019.

\bibitem[{Springel}(2005)]{springel2005cosmological}
Volker {Springel}.
\newblock {The cosmological simulation code GADGET-2}.
\newblock \emph{Monthly Notices of the Royal Astronomical Society},
  364\penalty0 (4):\penalty0 1105--1134, December 2005.
\newblock \doi{10.1111/j.1365-2966.2005.09655.x}.

\bibitem[{Schneider} et~al.(1992){Schneider}, {Ehlers}, and
  {Falco}]{Schneider1992gravitational}
Peter {Schneider}, J{\"u}rgen {Ehlers}, and Emilio~E. {Falco}.
\newblock \emph{{Gravitational Lenses}}.
\newblock 1992.
\newblock \doi{10.1007/978-3-662-03758-4}.

\bibitem[{Lu} and {Haiman}(2021)]{Lu2021impact}
Tianhuan {Lu} and Zolt{\'a}n {Haiman}.
\newblock {The impact of baryons on cosmological inference from weak lensing
  statistics}.
\newblock \emph{Monthly Notices of the Royal Astronomical Society},
  506\penalty0 (3):\penalty0 3406--3417, September 2021.
\newblock \doi{10.1093/mnras/stab1978}.

\bibitem[{Petri}(2016)]{Petri2016mocking}
A.~{Petri}.
\newblock {Mocking the weak lensing universe: The LensTools Python computing
  package}.
\newblock \emph{Astronomy and Computing}, 17:\penalty0 73--79, October 2016.
\newblock \doi{10.1016/j.ascom.2016.06.001}.

\bibitem[Martinet et~al.(2018)Martinet, Schneider, Hildebrandt, Shan, Asgari,
  Dietrich, Harnois-D{\'e}raps, Erben, Grado, Heymans,
  et~al.]{martinet2018kids}
Nicolas Martinet, Peter Schneider, Hendrik Hildebrandt, HuanYuan Shan, Marika
  Asgari, J{\"o}rg~P Dietrich, Joachim Harnois-D{\'e}raps, Thomas Erben,
  Aniello Grado, Catherine Heymans, et~al.
\newblock Kids-450: cosmological constraints from weak-lensing peak
  statistics--ii: Inference from shear peaks using n-body simulations.
\newblock \emph{Monthly Notices of the Royal Astronomical Society},
  474\penalty0 (1):\penalty0 712--730, 2018.

\bibitem[Harnois-D{\'e}raps et~al.(2021)Harnois-D{\'e}raps, Martinet, Castro,
  Dolag, Giblin, Heymans, Hildebrandt, and Xia]{harnois2021cosmic}
Joachim Harnois-D{\'e}raps, Nicolas Martinet, Tiago Castro, Klaus Dolag,
  Benjamin Giblin, Catherine Heymans, Hendrik Hildebrandt, and Qianli Xia.
\newblock Cosmic shear cosmology beyond two-point statistics: a combined peak
  count and correlation function analysis of des-y1.
\newblock \emph{Monthly Notices of the Royal Astronomical Society},
  506\penalty0 (2):\penalty0 1623--1650, 2021.

\bibitem[Z{\"u}rcher et~al.(2022)Z{\"u}rcher, Fluri, Sgier, Kacprzak, Gatti,
  Doux, Whiteway, Refregier, Chang, Jeffrey, et~al.]{zurcher2022dark}
Dominik Z{\"u}rcher, Janis Fluri, Rapha{\"e}l Sgier, Tomasz Kacprzak, Marco
  Gatti, Cyrille Doux, Lorne Whiteway, Alexandre Refregier, Chihway Chang,
  Niall Jeffrey, et~al.
\newblock Dark energy survey year 3 results: Cosmology with peaks using an
  emulator approach.
\newblock \emph{Monthly Notices of the Royal Astronomical Society},
  511\penalty0 (2):\penalty0 2075--2104, 2022.

\bibitem[Liu et~al.(2023)Liu, Yuan, Pan, Zhang, Wang, and
  Fan]{liu2023cosmological}
Xiangkun Liu, Shuo Yuan, Chuzhong Pan, Tianyu Zhang, Qiao Wang, and Zuhui Fan.
\newblock Cosmological studies from hsc-ssp tomographic weak-lensing peak
  abundances.
\newblock \emph{Monthly Notices of the Royal Astronomical Society},
  519\penalty0 (1):\penalty0 594--612, 2023.

\bibitem[Mallat(2012)]{mallat2012group}
St{\'e}phane Mallat.
\newblock Group invariant scattering.
\newblock \emph{Communications on Pure and Applied Mathematics}, 65\penalty0
  (10):\penalty0 1331--1398, 2012.

\bibitem[Cheng and M{\'e}nard(2021)]{cheng2021weak}
Sihao Cheng and Brice M{\'e}nard.
\newblock Weak lensing scattering transform: dark energy and neutrino mass
  sensitivity.
\newblock \emph{Monthly Notices of the Royal Astronomical Society},
  507\penalty0 (1):\penalty0 1012--1020, 2021.

\bibitem[Valogiannis and Dvorkin(2022)]{valogiannis2022towards}
Georgios Valogiannis and Cora Dvorkin.
\newblock Towards an optimal estimation of cosmological parameters with the
  wavelet scattering transform.
\newblock \emph{Physical Review D}, 105\penalty0 (10):\penalty0 103534, 2022.

\end{thebibliography}

%%%%%%%%%%%%  Supplementary Figures  %%%%%%%%%%%%
%\clearpage

%%%%%%%%%%%%%%%%   End   %%%%%%%%%%%%%%%%
%\end{multicols}  % Method B for two-column formatting (doesn't play well with line numbers), comment out if using method A
\end{document}